\documentclass[a4paper,12pt]{book}

\usepackage[utf8x]{inputenc}
\usepackage[greek,english]{babel}
\usepackage{amsmath}
\usepackage{amssymb}
\usepackage{latexsym}
\usepackage{graphicx}
\usepackage{epsfig}
\usepackage{color}

\usepackage[a4paper]{geometry}
\allowdisplaybreaks[1] %an xreiastei kanei split to align se duo selides 
\usepackage{afterpage}  

\newcommand\blankpage{%   
    \null
    \thispagestyle{empty}%
    \addtocounter{page}{-1}%
    \newpage}

\usepackage{fancyhdr}

\usepackage[colorlinks=true,
            linkcolor=blue,
            urlcolor=cyan,
            citecolor=red]{hyperref}

\def\beq{\begin{equation}}
\def\eeq{\end{equation}}
\def\bea{\begin{eqnarray}}
\def\eea{\end{eqnarray}}

\def\a{\alpha}
\def\b{\beta}
\def\r{\rho}
\def\d{\delta}
\def\m{\mu}
\def\n{\nu}
\def\h{\eta}
\def\s{\sigma}
\def\e{\varepsilon}
\def\l{\lambda}
\def\g{\gamma}
\def\G{\Gamma}
\def\F{\Phi}
\def\p{\pi}
\def\O{\mathcal{O}}
\def\hh{\tilde{h}}

\def\da{\partial^2}
\def\aa{\partial}
\newcommand{\an[1]}{\partial_#1}
\def\2{\;\;}
\def\4{\;\;\;\;}

\begin{document}

\begin{titlepage}
\centering

\includegraphics[width=1cm,height=2cm]{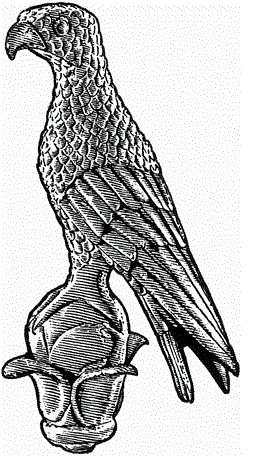}\\ [0.5cm]

 {\bf UNIVERSITY OF IOANNINA}\\
 %FACULTY
\bf SCHOOL OF NATURAL SCIENCES\\
  PHYSICS DEPARTMENT \\[3.5cm]

{\LARGE \bf  Gravitoelectromagnetism:\\[0.2cm]}
{\Large \bf Basic principles, novel approaches and their application to Electromagnetism } \\[3.5cm]

\bf {Athanasios Bakopoulos } \\[2.5cm]

MASTER THESIS\\[2.5cm]

\vfill

% Bottom of the page
{ IOANNINA 2016}

 \afterpage{\blankpage} 
\end{titlepage}

%%%%%%%%%%%%%%%%%%%%%%%%%%%%%%%%%%%%%%%%%%%%%%%%%%%%%%%
%%%%%%                                  %%%%%%%%%%%%%%%
%%%%%%           efxaristies            %%%%%%%%%%%%%%%
%%%%%%                                  %%%%%%%%%%%%%%%
%%%%%%%%%%%%%%%%%%%%%%%%%%%%%%%%%%%%%%%%%%%%%%%%%%%%%%%

\clearpage
\thispagestyle{empty}

\vspace*{4cm}
\hfill To my family
\afterpage{\blankpage}

\clearpage
\thispagestyle{empty}
\vspace*{4cm}
  
\begin{center}
{\bf Acknowledgements}\\[1cm]
\end{center}
First of all, I would like to thank my supervisor, Professor  Panagiota Kanti, for giving me the opportunity to work with her on this fascinating subject and also for her guidance, help and advice. 

I would also like to thank Professor K. Tamvakis and Professor L. Perivolaro- poulos for their advices. 

Last but not least, many thanks to my family and friends for their love and support.

\afterpage{\blankpage}

%%%%%%%%%%%%%%%%%%%%%%%%%%%%%%%%%%%%%%%%%%%%%%%%%%%%%%%
%%%%%%                                  %%%%%%%%%%%%%%%
%%%%%%           abstract             %%%%%%%%%%%%%%%
%%%%%%                                  %%%%%%%%%%%%%%%
%%%%%%%%%%%%%%%%%%%%%%%%%%%%%%%%%%%%%%%%%%%%%%%%%%%%%%%

\clearpage
\thispagestyle{empty}
\vspace*{4cm}
    
\begin{center}
{\bf Abstract}\\[1cm]
\end{center}

This work is focused on the theory of Gravitoelectromagnetism (GEM). In the first part of this work we present a brief review of gravitoelectromagnetism, we locate and discuss all the problems which appear in this approach. We also try to avoid these problems by proposing new approaches in which we use the additional degrees of freedom of the gravitational field. In the second part of this work, we review our previous work regarding the construction of a tensorial theory, using the formalism of General Relativity, which aims to describe the true electromagnetism. We also extend this theory in order to make it more realistic. Finally in the third part of this work, we investigate the 
existence of gravitational invariants similar to the electromagnetic ones. 

\afterpage{\blankpage}

\clearpage
\thispagestyle{empty}

%\afterpage{\blankpage} 
\clearpage
\thispagestyle{empty}
%%%%%%%%%%%%%%%%%%%%%%%%%%%%%%%%%%%%%%%%%%%%%%%%%%%%%%%
%%%%%%                                  %%%%%%%%%%%%%%%
%%%%%%           contents               %%%%%%%%%%%%%%%
%%%%%%                                  %%%%%%%%%%%%%%%
%%%%%%%%%%%%%%%%%%%%%%%%%%%%%%%%%%%%%%%%%%%%%%%%%%%%%%%

\pagestyle{fancy}
\fancyhead{}
\fancyhead[LO]{}
\fancyhead[LE]{Contents}

\tableofcontents

\clearpage
\thispagestyle{empty}

\latintext
\chapter*{Preface}
\label{1}
\addcontentsline{toc}{chapter}{\nameref{1}}
\fancyhead{}
\fancyhead[LO]{}
\fancyhead[LE]{Preface}

A problem which existed from the ancient times was the explanation of the planetary orbits.  In the 17th century this problem  was  theoretically explained by Newton's theory for Gravitation. Although Newton's theory seemed to explain everything regarding the planetary orbits, it could not explain a few anomalies in Mercuries orbit. In order to explained these anomalies, many new theories  were proposed and one of them is the gravitoelectromagnetism.

Gravitoelectromagnetism (GEM) is an approach in which the gravitation field is described using the formulation and the terminology of electromagnetism.  It is based on the profound analogy between the Newton's law for gravitation and Coulomb's law for electricity. Gravitoelectromagnetism has a long history, which starts in 1832, when Faraday  --who was convinced that the gravitational field unifies with the electromagnetic one--  conducted some experiments in order to detect if the gravitational field can induct electric current in a circuit, and which continues until our days. 

For about two centuries many papers and new approximations of Gravitoelectromagnetism have been proposed. However, during the last years, the majority of the scientific articles focusing on  gravitoelectromagnetism are based on Einstein's General Relativity and especially on the work of Lense and Thirring. Lense and Thirring   showed that in the context of General Relativity a rotating massive body can create a gravitomagnetic field. This effect is known as the Lense-Thirring effect.  These works are split in two categories. The first one uses the linearised form of General Relativity field equations in the weak field approximation, while the other uses the decomposition of the Weyl tensor into gravitoelectromagnetic parts. 

In this project we follow the first category, and work in the linear  approximation. The main achievements of this approach have been reviewed by B. Mashhoon. However,  the last years many researchers have noticed some problems in this  approach. Several scientific articles, including  one of ours, have been published in which solutions to the above problems have been proposed. Our objective for this project is to review  our previous work and to propose new solutions to the problems which have been reported.

Another topic, also covered in our published work, is an attempt to describe true electromagnetism using the formalism of General Relativity. This attempt is based on the results of our work regarding gravitoelectromagnetism. We should mention that our goal is not to replace electromagnetism with another theory but to see how far the analogy  with GEM extends and where and why it breaks down. In this thesis we continue our work  in this subject by proposing new approaches. 

In the last chapter of this thesis, we also cover  the search for invariant quantities similar to the electromagnetic ones, in the linear approach of gravitoelectromagnetism. This work is still in progress and we only present some of our results. We hope that this work would contribute  to a better understanding of gravitoelectromagnetism and, maybe,  lead to a lagrangian formulation.

The outline of this thesis is as follows: in chapter \ref{2} we present a historical overview  of gravitoelectromagnetism and a short introduction to General Relativity and its linear approximation. In chapter \ref{3} we give a brief review of gravitoelectromagnetism based on Mashhoon's papers and our previous work. In this chapter we locate and discuss all the problems which appear in this approach. In chapter \ref{4} we try to solve the problems which appear in gravitoelectromagnetism by proposing new approaches in which we use the additional degrees of freedom of the gravitational field. In chapter \ref{5} we review our previous work regarding the construction of a tensorial theory, using the formalism of General Relativity, which aims to describe the true electromagnetism. We also extend this theory in order to make it more realistic. In chapter \ref{6} we investigate the 
existence of gravitational invariants similar to the electromagnetic ones. Finally in chapter \ref{7} we present our conclusions.   

%%%%%%%%%%%%%%%%%%%%%%%%%%%%%%%%%%%%%%%%%%%%%%%%%%%%%%%%%%%%%%%%%%%%
%
%%%%%%%%%%%%%%%%%%%%%%%%%%%%%%%%%%%%%%%%%%%%%%%%%%%%%%%%%%%%%%%%%%%%%

\chapter{Introduction}\label{2}

\fancyhead{}
\fancyhead[LO]{\slshape\nouppercase{\rightmark}}
\fancyhead[LE]{\slshape\nouppercase{\leftmark}}
\fancyfoot{}
\fancyfoot[CE,CO]{\thepage}
%\fancyfoot[LO,LE]{Chapter \thechapter}
%\fancyfoot[RO,RE]{Section \thesection} 

\section{A historical review  of Gravitoelectromagnetism}\label{A000}

Gravitoelectromagnetism is  a theory in which  the main goal is to describe  gravity using the formulation of electromagnetism. Gravitoelectromagnetism as a theory has a long history. 

In 1832 Faraday conducted a series of experiments in order to detect the induction of electric current in a coil falling through the gravitational field of the Earth \cite{far}. 33 years later, Maxwell, motivated from the analogy between the Newton's law of gravitation and Coulomb's law of electricity, tried to develop a theory for gravitation using the formulation of electromagnetism \cite{maxw}. At the same time, Holzmuller \cite{hol} and Tisserand  \cite{tis} presented their attempts  to explain the excess advance in the perihelion of Mercury in terms  of a gravitomagnetic field generated by the Sun. In 1893, Heaviside \cite{hev} developed a full set of Maxwell-like equations for the gravitational field. Heaviside's field equations  in the vacuum can give wave solutions which propagate with the speed of light. These attempts were very promising for the explanation of the excess advance in the perihelion of Mercury, however, in 1900, Lorentz \cite{lor} proved that a gravitomagnetic force would be too weak to explain this phenomenon. 

In 1915, Einstein published the General Theory of Relativity \cite{Einst}. General Relativity is a modern theory of gravitation which has extended Newton's theory. General Relativity has, also, provided an explanation to the problem of the  perihelion of Mercury. A few years later, Lense and Thirring, using General Relativity, showed that a rotating massive body can create a gravitomagnetic field \cite{LT}; this is known as the Lense-Thirring effect. Thirring also proved that geodesics equation can be expressed in terms of a gravitoelectric and a gravitomagnetic field in the same way as the Lorentz force for electromagnetism \cite{thi}. 

In the 1950's, Matte \cite{mat}, Bel \cite{bel} and Debever \cite{deb} explored the analogy of gravity and electromagnetism based on the decomposition of the Weyl tensor into traceless tensors in the weak approximation limit. In the 1960's, Forward \cite{forw1} extended the above model and obtained  a Maxwell structure for the perturbation; he also proposed experiments to detect these fields \cite{forw2}. In this decade, a new approach appeared, presented by Scott  \cite{sco}, in which the gravitational field is expressed by a vector field in strict analogy to the electromagnetic field. Similar vector theories have been proposed by Coster and Shepanski \cite{cost} and by Schwebel \cite{sch}. In the Seventies, another vector theory was presented by Spieweck \cite{spi} in which a gravitomagnetic vector potential was introduced. Also, in order to explain the strange behavior of astrophysical objects under extreme physical conditions, new vector approaches of gravitoelectromagnetism were published \cite{maj}\cite{bri}\cite{whit}. The above approaches have many applications in Astrophysics.

During the last thirty years many approaches of gravitoelectromagnetism, based on General Relativity,  appeared in the literature  \cite{gem}\cite{Harris}. We can say that the gravitoelectromagnetic formulations can be split into two categories. The first, which is the one we use here, is an approach based in the linearised form of General Relativity field equations around a massive rotating body. A review of the linear approach has been  published by Mashhoon  \cite{Mashhoon} and it is the starting point for many papers. The second is an approach based on tidal tensors and on the decomposition of the Weyl tensor into gravitomagnetic and gravitoelectric parts. This approach has been extensively reviewed by Costa, Natario and Herdeiro \cite{Costa}\cite{Natario}. However, new approaches are still appearing in the literature, such as the formulation based on group theory by Ramos \cite{ram}  or the Langrangian formulation of gravitoelectromagnetism published by Ramos, Montigny and Khanna \cite{ram2}. Many interesting phenomena are associated to gravitoelectromagnetism: Ruggiero and Tartaglia have  published a review  of the gravitoelectromagnetic effects \cite{tart} which also contains  references to experimental attempts to detect them.

In the following two sections we give a brief introduction to General Relativity and to the linear approach. We consider   this   specifically important in order to give the context in which our work is based  and to provide the reader with the information needed to understand better our project.

\section{General Relativity}\label{A0}

The General   Theory of Relativity is the modern theory of gravitation. It was published by A. Einstein in 1915 \cite{Einst}, and is a theory that connects the gravitational force with the space-time geometry. 

The geometry in General Relativity is described by the metric tensor $g_{\m\n}$ which is a second rank symmetric tensor. The metric tensor is appearing in the well-known ``first fundamental form" of the Differential Geometry which is also known as the line element
\begin{equation}
ds^2=g_{\m\n}dx^\m dx^\n,
\end{equation}  
where $x^\mu=(ct, \vec{x})$. The above equation is a generalized form of the well-known line element of the Special Theory of Relativity \cite{Einst2}
\begin{equation}
ds^2=\h_{\m\n}dx^\m dx^\n,
\end{equation} 
where $\h_{\m\n}$ is the metric tensor of the flat  Minkowski space-time\footnote{Throughout this work, we will use the $(+1,-1,-1,-1)$ signature 
for the Minkowski tensor $\eta_{\mu\nu}$.}.

The metric tensor is also related with the curvature of the space which is described by the Riemann's --curvature-- tensor
\begin{equation}
R^\r_{\2\,\s\m\n}= \aa_\m  \G^{\r}_{\2\,\n\s} - \aa_\n  \G^{\r}_{\2\,\m\s} + \G^{\r}_{\2\,\m\l}\G^{\l}_{\2\,\n\s} -  \G^{\r}_{\2\,\n\l}\G^{\l}_{\2\,\m\s}\,,
\end{equation} 
where $\G^{\r}_{\2\,\m\n}$ is the Christoffel's symbols   given by the following combination of  the metric tensor and its derivatives
%%%%
\begin{equation}
\G^{\r}_{\2\,\m\n}=\frac{1}{2}g^{\l\r} \left( \aa_\m g_{\l\n} + \aa_\n g_{\l\m} - \aa_\l g_{\m\n} \right).
\end{equation}
Two useful geometrical quantities for the General Relativity, which can be extracted from the Riemann  tensor, are the Ricci  tensor which is defined as 
%%%
\begin{equation}\label{ricc}
R_{\m\n}=R^\r_{\2\,\m\r\n}= \aa_\r  \G^{\r}_{\2\,\m\n} - \aa_\m  \G^{\r}_{\2\,\r\n} + \G^{\r}_{\2\,\r\l}\G^{\l}_{\2\,\m\n} -  \G^{\r}_{\2\,\m\l}\G^{\l}_{\2\,\r\n}\,,
\end{equation}
%%%%%%
and its trace  defined as
%%%%%
\begin{equation}\label{riccsca}
R=g^{\m\n}R_{\m\n}.
\end{equation}
The trace of the Ricci tensor   is also known as the  Ricci scalar.

The field equations of the General Relativity are 
%%%%
\begin{equation}\label{fieldeq}
G_{\m\n}=\frac{8 \p G}{c^4} T_{\m\n},
\end{equation}
where the tensor $G_{\m\n}$ is   called Einstein's tensor and  is defined as 
\begin{equation}\label{eins}
G_{\m\n}=R_{\m\n}-\frac{1}{2}g_{\m\n}R.
\end{equation}
The tensor $T_{\m\n}$ is the so-called energy-momentum tensor which describes the distribution of mass and energy in the region of   space where we solve the field  equations.

In General Relativity the equations of motion of a particle which moves only under the gravitational force is given by the geodesics equation %%%%%%%%%%%%%%%%%%%
\begin{equation} \label{geodesics-GEM}
\frac{d^2 x^\rho}{ds^2} + \Gamma^\rho_{\mu\nu}\,\frac{dx^\mu}{ds}
\frac{dx^\nu}{ds}=0. 
\end{equation}
%%%%%%%%%%%%%%%%%%% 
As in the Newton's theory of gravity the motion is independent of the particle's mass.

\section{The linear approach}\label{A1}

Einstein's field equations are non linear and thus, quite often, difficult to solve. A usual technique is to adopt the linear approach valid in the limit of the weak gravitational field. In this the metric tensor may be written as 
%%%%%%%%%%%%
\begin{equation}
g_{\mu\nu} (x^\mu) =\eta_{\mu\nu} + h_{\mu\nu}(x^\mu)\,,
\label{metric}
\end{equation}
%%%%%%%%%%%%%%
where $h_{\mu\nu}$ are small perturbations over the Minkowski space-time. In the context of General Relativity, the metric perturbations
$h_{\mu\nu}$  are sourced by gravitating bodies and  obey the
inequality $|h_{\mu\nu}|\ll 1$. As a result a linear-approximation
analysis may be followed. Here we will briefly review the corresponding formalism and we will present the equations of General Relativity 
in the linear-order approximation (for a more detailed analysis, see for
example \cite{Landau}). If we use Eq. (\ref{metric}) and keep only terms 
linear in the perturbation $h_{\mu\nu}$, we easily find that the Christoffel
symbols take the form
%%%%%%%%%%%%%%%
\begin{equation}\label{Christoffel}
\Gamma^{\alpha}_{\mu\nu}=\frac{1}{2}\,\eta^{\alpha\rho}\left(
h_{\mu\rho,\nu}+h_{\nu\rho,\mu}-h_{\mu\nu,\rho}\right).
\end{equation}
%%%%%%%%%%%%%%%%%
Using the  above equation, the Ricci tensor (Eq. (\ref{ricc})) assumes the form
%%%%%%%%%%%%%%%%%%
\begin{equation}\label{RicciT}
R_{\mu\nu}=\frac{1}{2}\left({h^\rho}_{\mu,\nu\rho}+
{h^\rho}_{\nu,\mu\rho}-\da h_{\mu\nu} - h_{,\mu\nu}\right),
\end{equation}
%%%%%%%%%%%%%%%
where $\da=\h^{\mu\nu}\aa_\m\aa_\n$. In the linear approximation the tensor indices are raised and lowered by
the Minkowski metric $\eta_{\mu\nu}$. In the above equation, we have also defined
the trace of the metric perturbations $h=\h^{\m\n}h_{\m\n}$. Correspondingly, the trace of the Ricci
tensor (Eq. (\ref{riccsca})) is found to be
%%%%%%%%%%%%%%%
\begin{equation}\label{RicciS}
R={h^{\m\n}}_{,\m\n} - \da h\,.
\end{equation}
%%%%%%%%%%%%%%%
If we combine the Ricci tensor and its trace, the Einstein tensor (Eq. (\ref{eins})) takes the form
%%%%%%%%%%%%%
\begin{equation}\label{Einstein}
G_{\m\n}=\frac{1}{2}\left({h^\a}_{\m,\n\a}+ {h^\a}_{\n,\m\a}-
\da h_{\m\n}-h_{,\m\n}-\h_{\m\n}\,{h^{\a\b}}_{,\a\b}+\h_{\m\n}\,\da h \right).
\end{equation}
%%%%%%%%%%%%%%%
Usually in gravity,  in the linear approximation, one may define the new perturbations $\hh_{\m\n}$ in terms of the original perturbations $h_{\m\n}$:
%%%%%%%%%%%%%%%
\begin{equation}\label{newh}
\hh_{\m\n}=h_{\m\n} - \frac{1}{2}\,\h_{\m\n}\,h\,.
\end{equation}
%%%%%%%%%%%%%%%
In terms of the new perturbations  the Einstein tensor simplifies to:
%%%%%%%%%%%%%%%%%
\begin{equation}\label{Einstein_full}
G_{\m\n}=\frac{1}{2}\left({\hh^{\a}}_{\2\m,\n\a}+{\hh^{\a}}_{\2\n,\m\a}-
\da \hh_{\m\n}-\h_{\m\n}\,{\hh^{\a\b}}_{\4,\a\b}\right).
\end{equation}
%%%%%%%%%%%%%%%%%%%

The above tensor satisfies Einstein's field equations Eq. (\ref{fieldeq}) which now are written as
%%%%%%%%%%%%%%%%
\begin{equation}\label{field_eqs_full}
{\hh^{\a}}_{\2\m,\n\a}+{\hh^{\a}}_{\2\n,\m\a}-
\da \hh_{\m\n}-\h_{\m\n}\,{\hh^{\a\b}}_{\4,\a\b}=2k\,T_{\mu\nu}\,.
\end{equation}
The value of the proportionality constant $k$   will be $\frac{8 \p G}{c^4}$, as usual,  except in chapter \ref{5} where it will be different. If we make a coordinate transformation of the form $x^\m\rightarrow x'^\m = x^\m - \g^\m$, we can easily see that the metric perturbations transform as $h_{\m\n} \rightarrow h'_{\m\n} = h_{\m\n} + \g_{\m,\n} + \g_{\n,\m} $. This transformation leaves Einstein tensor unchanged and hence the field equation. Transformations of this form, which do not alter  observable quantities, are called {\it gauge transformations}.  When we solve a problem, it is often convenient  to reduce the gauge freedom by imposing constraints to the perturbations $h_{\m\n}$. These constraints are called  {\it gauge conditions}. Usually, in the linear approximation, we use the so-called {\it transverse} gauge condition ${\hh^{\m\n}}_{\4,\n}=0$. Under the imposition of the transverse gauge condition the Einstein  tensor takes the simple form
%%%%%%%%%%%%%%%
\begin{equation}\label{Einstein_new}
G_{\m\n}=-\frac{1}{2} \da \hh_{\m\n},
\end{equation}
%%%%%%%%%%%%%%%
while the field equations are
%%%%%%%%%%%%%%%%
\begin{equation}\label{field_eqs}
\da \hh_{\m\n}=-2k\,T_{\mu\nu}\,.
\end{equation}
%%%%%%%%%%%%%%%
If we multiply Eqs. (\ref{field_eqs}) with $\h^{\m\n}$, we get $ \da \hh=-2k\,T$ or $\da h=2k\,T$ and we can write the field equations as:
%%%%%%%%
\beq\label{field_alt}
\da h_{\m\n}=-2k\,\left(T_{\mu\nu}-\frac{1}{2}\h_{\m\n}T\right)\,.
\eeq
%%%%%%%%%%%%%%
In the vacuum we have $T_{\m\n}=0\,$ and the field equations take the form:
\begin{equation}\label{field_vac}
\da \hh_{\m\n}=0 \4\4 or \4\4 \da h_{\m\n}=0\, .
\end{equation}
In what follows we will assume that the distribution of energy in the
system is described by the expression $T_{\mu\nu}=\rho\,u_\mu u_\nu$,
where $\rho$ is the mass density and $u^\mu=(u^0,u^i)=(c,\vec{u})$ is the
velocity of the source. If we define the mass current density as $j_\m = \r \,u_\m$, the energy-momentum tensor takes the form $T_{\mu\nu}= j_\mu u_\nu$.

%%%%%%%%%%%%%%%%%%%%%%%%%%%%%%%%%%%%%%%%%%%%%%%%%%%%%%%%%%%%%%%%%%%%%

\chapter{GravitoElectroMagnetism}\label{3}

\section{The traditional ansatz for the perturbations}\label{A}
In the context of the theory of gravitoelectromagnetism, Eqs. (\ref{field_eqs})
are Einstein's linearised gravitational field equations, with $k=8\pi G/c^4$, under the imposition of the transverse gauge condition.
The components of the metric perturbations $\tilde h_{\mu\nu}$ have
the form \cite{Mashhoon}
%%%%%%%%%%%%
\beq 
\tilde h_{00}=\frac{4\Phi}{c^2}\,, \qquad \tilde h_{0i}=\frac{2A_i}{c^2}\,,
\qquad \tilde h_{ij}=\,\mathcal{O}(c^{-4})\, \label{case1}
\eeq
%%%%%%%%%%%
or 
\begin{equation}
\tilde h_{0\m}= \frac{4}{c^2} A_{\m}\,,  \qquad \tilde h_{ij}=\,\mathcal{O}(c^{-4})\,
\end{equation}
where $A^{\m}=\left(\F\,,\frac{\vec{A}}{2}\,\right)\,$ is the gravitoelectromagnetic 4-potential.
If we contract
Eq. (\ref{newh}) by $\eta^{\mu\nu}$ we find $h=-\tilde{h}$; and we can 
write the inverse relation between the original and the new perturbations as 
%%%%%%%%%%%%%%%%%
\beq
h_{\mu\nu}= \tilde h_{\mu\nu} -\frac{1}{2}\,\eta_{\mu\nu}\,\tilde{h}\,.
\label{inversehnew}
\eeq
%%%%%%%%%%%%%%%%%%%
Then, using the definition (\ref{metric}), the spacetime line-element
takes the form 
%%%%%%%%%%%%%%%%%%
\begin{align}\label{line-element-GEM}
ds^2=c^2\left(1+\frac{2\Phi}{c^2}\right) dt^2-\frac{4}{c}\,(\vec{A} \cdot d\vec{x})\,dt
- \left(1-\frac{2\Phi}{c^2}\right) \delta_{ij} dx^i dx^j.
\end{align}
The perturbations  are thus expressed in terms of the so-called gravitoelectromagnetic potentials,  a scalar $\Phi(x^\mu)$ and a vector potential
$\vec{A}(x^\mu)$. The $\tilde h_{00}$ component yields
the Newtonian potential $\Phi$, while the $\tilde h_{0i}$ component is associated
to the ``vector" potential $\vec{A}$ generated by a rotating massive body; the $\tilde h_{ij}$
component of the metric perturbations is usually assumed to be negligible due to the suppression of the
corresponding source by a $1/c^4$ factor.  

The transverse gauge, $\hh^{\m\n}_{\4,\n}=0$, gives:
%%%%%%%%%%%%%%%%%%%%%%%%%%%%%%%%%%%
\begin{itemize}

\item For $\m=0$, the analogue to the Lorentz gauge condition of the electromagnetism :
\begin{equation}\label{lor567}
\frac{1}{c}\,\aa_t \F + \aa_i \left(\frac{A^i}{2}  \right)=0 \4\4or\4\4 A^{\,\m}_{\,\,\,\, ,\m}=0.
\end{equation}

\item And for $\m=i$, an additional condition that connects the $\tilde h_{ij}$
component with the time derivative of the vector potential $\vec{A}$:

\begin{equation}\label{add_gauge1}
\aa_j \hh^{ij} \,=\,-\,\aa_t\,A^i.\,\,
\end{equation}

\end{itemize}
%%%%%%%%%%%%%%%%%%%%%%
Due to the tensorial structure of gravity, from the gauge condition we get three more equations, Eq. (\ref{add_gauge1}),  that do not appear  in the true electromagnetism. These equations  were ignored in many scientific articles  because they involve $\O(c^{-4})$ terms \cite{Mashhoon}\cite{Mash2} and  lead to static vector potential $\aa_t\,\vec{A}=0$ in some others \cite{Bakop}\cite{Costa}. Here we are following   Mashhoon's papers \cite{Mashhoon} and we are ignoring the presence of these equations. In the following sections we will discuss further the presence and   importance of  Eq (\ref{add_gauge1}).

The field equations are:
%%%%%%%
\begin{itemize}

\item For $\m=0 $ and $\n=0\,$:
\begin{equation}\label{poison_1}
\da \F\,=\,4 \p  G \, \r,
\end{equation}

\item And for $\m=0 $ and $\n=i\,$:
\begin{equation}\label{dianysm_1}
\da \left(\frac{A^i}{2} \right)\,=\,\frac{4\p G}{c}\,j^i,
\end{equation}
\end{itemize}
%%%%
where we have ignored all the equations involving $\O(c^{-4})$ terms.  

In analogy with electromagnetism, we can define the GEM fields $\vec{E}$ and $\vec{B}$
in terms of the GEM potentials \cite{Mashhoon}\cite{Bakop}
%%%%%%%%%%%%%%%
\beq
\vec{E} \equiv -\frac{1}{c}\,\partial_t \left(\frac{\vec{A}}{2}\right) -\vec{\nabla} \Phi\,, 
\qquad \vec{B} \equiv \vec{\nabla} \times \left(\frac{\vec{A}}{2}\right)\,. \label{EB_GEM}
\eeq
%%%%%%%%%%%%%%%
According to \cite{Mashhoon}, if we use vector notation, Eqs. (\ref{poison_1}) and (\ref{dianysm_1}) reduce to two Maxwell-like equations for the Gravitoelectromagnetic field
%%% 
%%%%%%%%%%%%%%
\beq
\vec{\nabla} \cdot \vec{E} = 4\pi G \rho\,, \qquad 
\vec{\nabla} \times \vec{B}=\frac{1}{c}\,\partial_t \vec{E} + 
\frac{4\pi G}{c}\,\vec{j}\,. \label{FinalGEM_1.1}
\eeq
%%%%%%%%

The definitions of the fields Eqs. (\ref{EB_GEM}) reduce to the remaining  Maxwell  equations 
\begin{equation}\label{FinalGEM_1.2}
\vec{\nabla} \times \vec{E}=\frac{1}{c}\,\partial_t \vec{B}, \4\4 \vec{\nabla} \cdot \vec{B} =0.
\end{equation}
A more detailed analysis, that will be presented in the following section, will give more information on how Eqs (\ref{poison_1}-\ref{dianysm_1}) actually reduce to a set of Maxwell-like equations.

Finally we can find the equation of motion of a test particle which propagates in the background (\ref{line-element-GEM}). Here, and in the following sections, we will keep all corrections in the equation of motion and we discuss their importance. We should also mention that we work in the non relativistic limit; this means that we can write
%%%%%%%%%%%%%%%%%
\beq
ds^2 = c^2dt^2-(dx^1)^2 -(dx^2)^2 -(dx^3)^2=
c^2 dt^2 \left(1-\frac{|\vec{u}|^2}{c^2}\right) \simeq c^2 dt^2\,.
\eeq
%%%%%%%%%%%%%%%%%%
In the non-relativistic limit, the spatial components of the geodesics equation  
%%%%%%%%%%%%%%%%%%%
\beq
\frac{d^2 x^\rho}{ds^2} + \Gamma^\rho_{\mu\nu}\,\frac{dx^\mu}{ds}
\frac{dx^\nu}{ds}=0 \nonumber
\eeq
%%%%%%%%%%%%%%%%%%%
take the  form
%%%%%%%%%%%%%%%%
\begin{equation}
\frac{d^2x^i}{dt^2}+ c^2\,\G^i_{00}+
2c \,\G^i_{0j}\,\frac{dx^j}{dt}+
\G^i_{kj}\,\frac{dx^k}{dt}\frac{dx^j}{dt} =0\,. \label{geodesics1-GEM}
\end{equation}
%%%%%%%%%%%%%%%
In the linear approximation, the Christoffel symbols are given by Eq. (\ref{Christoffel}). 
Reading the form of the initial perturbations $h_{\mu\nu}$ from the line-element
(\ref{line-element-GEM}), we find:
%%%%%%%%%%%%%%
\begin{eqnarray}
\G^i_{00} &=& 
%\frac{1}{2}\,\h^{ij}\left(\frac{2}{c}\, \aa_t h_{0j}-\aa_j h_{00}\right)=
-\frac{1}{c^2}\,\partial^i \Phi
+\frac{2}{c^3}\,\partial_t A^i\,, \\[2mm]
%%%%
\G^i_{0j} &=& 
%\frac{1}{2}\,\h^{ik}\left(\an[j]h_{0k}-\an[k]h_{0j}\right)=
\frac{1}{c^2}\,F_{j}^{\2\, i} - 
\,\delta^i_j \,\frac{1}{c^3} \partial_t \Phi, \\[2mm]
%%%%
\G^i_{kj} &=& 
%\frac{1}{2}\,\h^{il}\left(h_{lj,k}+h_{lk,j}-h_{jk,l}\right)=
-\frac{1}{c^2}\left(\delta^i_j \,\partial_k \Phi +
\delta^i_k \,\partial_j \Phi -\h_{kj} \,\partial^i \Phi \right).
\end{eqnarray}
%%%%%%%%%%%%%%%
In the second of the above equations we use the definition 
$F_{ij} \equiv \partial_i A_j -\partial_j A_i$. Substituting the above into
Eq. (\ref{geodesics1-GEM}) \cite{Mash3}\cite{Bakop}\cite{Costa}  we find
%%%
\begin{align}\label{lorentz_full1}
\ddot{x}^{\,i}=\,&E^i +\frac{2}{c}F^{ij}u_j+\frac{1}{c^2}\,\left(\,2u^i\aa_t \F +2u^i u^k \aa_k \F\,-u^k u_k \aa^i \F\,\right),
\end{align}
%%%%%%
which, if we are in the non relativistic limit and the scalar and the vector potential are stationary i.e. $\aa_t\,\F\,=\,0$ and $\aa_t\vec{A}\,=\,0$, reduces to the analogue of the Lorentz force law for   gravitation:
%%%%%
\begin{equation}\label{lorentz1}
\ddot{x}^{\,i}=\,E^i +\frac{2}{c}F^{ij}u_j\,,
\end{equation}
%%%%%%%
with ``electric charge" equal to $m$ and ``magnetic charge" equal  to $2m$. According to the literature the coefficient $2$ in the ``magnetic charge" is related to the spin of the gravitational field \cite{Mashhoon}. Actually this coefficient depends on the definition of the magnetic field and on the $h_{0i}$ component of the metric perturbations as well. From the definition of the magnetic field Eq. (\ref{EB_GEM}), we can see   that the $F_{ij}$ components are $F_{ij}=-2\e_{ijk}B_k$ and, as a result, the coefficient in front of the magnetic field is $4m$. In the above equation $\e_{ijk}$ is the Levi-Civita symbol.

With the traditional ansatz (Eq. (\ref{case1})) one can easily see the analogy 
between gravity and electromagnetism and, by solving the field  
equations, one may study many interesting phenomena \cite{LT}
\cite{Teys}. However, this ansatz ``works", i.e. it keeps the analogy with true electromagnetism, only if the 
potentials are  stationary and, as it is mentioned before, some times --through the gauge condition--  the ansatz 
dictates  that the vector 
potential $\vec{A}$ is time-independent. 
  If the potentials are time-dependent, the 
theory gives correct results, from the point of view of  General 
Relativity, but the analogy with the electromagnetism breaks 
down.
The truth is that, in reality, the 
majority of the gravitoelectromagnetic effects are stationary 
and well described by the traditional ansatz. But we are also 
interested in time-dependent effects and  we want to see if there is 
any way to preserve the analogy. In the literature  a number of  works have been 
found attempting to solve this problem 
\cite{Mashtime} but they are dealing only with special  cases. On 
the other hand, another big problem with 
gravitoelectromagnetism is that, even in the stationary case, 
we get the correct form of the Lorentz force law only in the 
non relativistic limit. It has been claimed \cite{Mashhoon} that we can keep 
the analogy ether with the field equations or  the Lorentz 
force but not   both with the same ansatz \cite{Costa}\footnote{See Section \ref{C}.}. If we 
choose an ansatz that fixes the Lorentz force law, we do not obtain 
the correct field equations \cite{Bakop}. In the following  sections 
we   present different ansatzes that attempt to fix the 
problems appearing  in the traditional one that we presented in this section.

\section{A detailed analysis of the traditional ansatz without the imposition of a gauge condition}\label{B}

As we saw in the previous section, the traditional ansatz  demonstrates the analogy with the true electromagnetism but it has some problems. In order to shed light to the problems of the traditional ansatz, in a previous work of ours \cite{Bakop}, we performed the analysis again without the imposition of a gauge condition. However, there is a difference between the ansatz we are using in this section and the one employed in section (\ref{A}). Here the $\hh_{ij}$ components of the metric perturbations $\hh_{\m\n}$ are set equal to zero \cite{Bakop}\cite{Costa}. Actually there is not a big difference since in the previous ansatz \cite{Mashhoon} these components were treated as null. However, as we will see in the following chapter, the presence of the $\hh_{ij}$ is in fact important for the solution of the problems which arise here.

  The ansatz we use in this section is
%%%%%%%%%%%%
\beq 
\tilde h_{00}=\frac{4\Phi}{c^2}\,, \qquad \tilde h_{0i}=-\frac{2A^i}{c^2}\,,
\qquad \tilde h_{ij}=0\,, \label{case2}
\eeq
%%%%%%%%%%%
and again the metric perturbations are expressed in terms of the so-called Gravitoelectromagnetic potentials.
%%%%%%

For the evaluation of the field equations we   need
the components of $\tilde h_{\mu\nu}$ in mixed form - these are:
%%%%%%%%%%%%
\beq 
\tilde h^{0}_{\ 0}=\frac{4\Phi}{c^2}\,, \qquad \tilde h^{0}_{\ i}=-\frac{2A^i}{c^2}\,,
\qquad \tilde h^{i}_{\ 0}=\frac{2A^i}{c^2}\,,\qquad \tilde h^{i}_{\ j}=0\,.
\eeq
%%%%%%%%%%%
The field equations, Eq. (\ref{field_eqs_full}),   for the above ansatz take the form
\begin{itemize}
\item For $\m=0$ and $\n=0$:
%%%%
\begin{equation}
\frac{\delta^{ij}}{c^2}\,\partial_i \partial_j \Phi =\frac{k}{2}\,\rho\,u_0\,u_0.
\label{Poisson0GEM}
\end{equation}
%%%%%%
\item For $\m=0$ and $\n=i$:
%%%%%
\begin{equation}
\frac{1}{c^2}\,\partial_i\left(\frac{1}{2}\,\partial_k A^k +\frac{1}{c}\,\partial_t \Phi\right)
-\frac{1}{2c^2}\,\delta^{kl}\,\partial_k \partial_l A^i = \frac{k}{2}\,\rho\,u_0\,u_i. 
\label{4Maxwell0GEM}
\end{equation}
%%%%%%
\item And finally for $\m=i$ and $\n=j$:
%%%%%
\begin{equation}
-\frac{1}{2c^3}\,\partial_t \left(\partial_i A^j +\partial_j A^i\right)
+ \delta_{ij}\left[\frac{1}{c^4}\,\partial_t^2 \Phi + 
\frac{1}{c^3}\,\partial_t (\partial_k A^k)\right] = \frac{k}{2}\,\rho\,u_i\,u_j\,.
\label{ExtraGEM}
\end{equation}
%%%%%%
\end{itemize}
%%%%%%%%%%%%%%%
As in the previous section, the constant $k$ is $8\pi G/c^4$. If we   use vector notation,  Eq. (\ref{Poisson0GEM})
readily takes the analogue of Poisson's law
%%%%%%%%%%%%%%%%
\beq
\nabla^2\,\Phi=4 \pi G \rho\,,
\label{Poisson_GEM}
\eeq
%%%%%%%%%%%%%%%
while Eq. (\ref{4Maxwell0GEM}) in turn can be rewritten as
%%%%%%%%%%%%%%%%
\beq
\vec{\nabla}\,\left[\vec{\nabla} \cdot \left(\frac{\vec{A}}{2}\right) + 
\frac{1}{c}\,\partial_t \Phi\right]
-\nabla^2 \left(\frac{\vec{A}}{2}\right) = \frac{4\pi G}{c}\,\rho \,\vec{u}\,.
\label{4MaxwellGEM}
\eeq
%%%%%%%%%%%%%%%%
Equations (\ref{Poisson_GEM}) and (\ref{4MaxwellGEM}) differ from (\ref{poison_1} - \ref{dianysm_1}) since here, we have not imposed the transverse gauge condition.
We can define again the GEM fields $\vec{E}$ and $\vec{B}$
in terms of the GEM potentials \cite{Mashhoon} \cite{Bakop}:
%%%%%%%%%%%%%%%
\beq
\vec{E} \equiv -\frac{1}{c}\,\partial_t \left(\frac{\vec{A}}{2}\right) -\vec{\nabla} \Phi\,, 
\qquad \vec{B} \equiv \vec{\nabla} \times \left(\frac{\vec{A}}{2}\right)\,. \label{EB_GEM2}
\eeq
%%%%%%%%%%%%%%%
Then, Eqs. (\ref{Poisson_GEM}) and (\ref{4MaxwellGEM}),using vector notation,
% along with the definitions (\ref{EB_GEM2}), reduce to a set of four Maxwell-like equations for the gravitoelectromagnetic field. The same equations as Eqs. (\ref{FinalGEM_1.1}) and (\ref{FinalGEM_1.2}):
take the form 
%%%%%%%%%%%%%%
\beq
\vec{\nabla} \cdot \vec{E} = 4\pi G \rho\,, \qquad 
\vec{\nabla} \times \vec{B}=\frac{1}{c}\,\partial_t \vec{E} + 
\frac{4\pi G}{c}\,\vec{j}\,, \label{FinalEqsGEM2.1}
\eeq
%%%%%%%%%%%%%
only if the vector  potential $\vec{A}$ has not  a time-dependence, i.e.
%%%%%%%%%%%%%%%%%%
\begin{equation}\label{vecnottime}
\aa_t\,\vec{A}\,=\,0\,.
\end{equation}
%%%%%%%%%%
The definitions (\ref{EB_GEM2}) then take the form
\begin{equation}\label{FinalEqsGEM2.2}
\vec{\nabla} \times \vec{E}=\frac{1}{c}\,\partial_t \vec{B}, \4\4 \vec{\nabla} \cdot \vec{B} =0.
\end{equation}

%The above equations may have the same form as in the previous section  but if one looks more carefully the Eqs. (\ref{Poisson0GEM}-\ref{ExtraGEM}), he will see that they reduce to Eqs. (\ref{FinalEqsGEM2.1}) 

%If we had  imposed the transverse gauge condition, we would get the same results (and that   was our starting point when we decided to work without a gauge condition). The transverse gauge condition gives the Lorentz gauge and Eq. (\ref{vecnottime}) as in the Eqs. (\ref{lor567} - \ref{add_gauge1}). As we have already mentioned   above the fact that the vector potential is static comes from the choice  of this specific ansatz and it does not depend on the gauge condition.
%
Here we can see that the choice we made for the form of the metric perturbations dictates that the vector potential is static (for different approaches regarding the role of this constraint in the context of gravitoelectromagnetism, see   \cite{Harris}\cite{Costa}\cite{Natario}\cite{Bakop}\cite{Braginsky}\cite{Pascual}). This does not fix the problems of the traditional ansatz; it actually makes them worst, however, it shows us that the time dependence of the vector potential depends on the ansatz. This is an important result, because in the previous section the constraint regarding the time dependence of the vector potential deduced from the gauge condition.  If we impose the transverse gauge condition we obtain the same results. Therefore, we can conclude that the only way to solve the problems regarding the time dependence of the vector potential  is to seek for new ansatzes. 
%
%and, through Eq. (\ref{add_gauge1}), on the  $\hh_{ij}$ components of the metric perturbations. %
% 
%
%In  Mashhoon's papers \cite{Mashhoon}\cite{Mash2}\cite{Mash3}\cite{Mashtime}, the spatial components of the gauge condition  ($\hh^{i\m}_{\,\,\,\,,\m}$) are not taken into account and the vector potential is assumed to be time dependent.  In the following chapter we will see  that, if we use the negligible components of the metric perturbation i.e. ($\hh_{ij}$), the problem of the time dependence of the vector potential is solved. In   Mashhoon's papers  the $\hh_{ij}$ components are not zero and, indeed, the vector potential may have time dependence, although nothing is said about the $\hh^{i\m}_{\,\,\,\,,\m}$ condition and its importance on the time dependence of the vector potential. What was  only mentioned here \cite{Mash2} is that this equations involve $\O(c^{-4})$ terms and for this reason they are negligible.
%
%Footnote end
 %Consequently, an alternative ansatz for the metric perturbations or the involvement of the $\hh_{ij}$ components on any ansatz   may fix  the problem. 

Now we can return to the linearised field equations. Their spatial component gives us a third set of constraints Eq. (\ref{ExtraGEM}).  The diagonal components of Eq. (\ref{ExtraGEM})
(i.e. for $i=j$) reduce to the relation
%%%%%%%%%%%%%%
\beq
\partial_t^2 \Phi =- \frac{\pi}{3} \rho\,|\vec{u}|^2\,,
\label{diagonal-GEM}
\eeq
%%%%%%%%%%%%%%
with $(u^1)^2=(u^2)^2=(u^3)^2$, and the off-diagonal ones (for $i \neq j$) give us 
%%%%%%%%%%%%%%
\beq
\partial_0 \left(\partial_i A^j +\partial_j A^i\right)= 8\pi G\rho 
\,\frac{u_i u_j}{c^2}\,. \label{off-diagonal-GEM}
\eeq
%%%%%%%%%%%%%%%
Therefore Eq. (\ref{diagonal-GEM})  demands that the distribution of sources
is isotropic (i.e. the current vector has the same   magnitude along all
three spatial directions); it also restricts the magnitude of $\partial_t^2 \Phi$,
which is a quantity that does not appear in the other two derived equations. The other
equation (\ref{off-diagonal-GEM}) imposes that the velocity of the source is
absolutely non-relativistic, $|\vec{u}| \ll c$: it is only then that the time 
variation of the vector potential is extremely small, due to the suppression
factor $1/c^2$ on its right-hand-side, and the consistency of the complete
set of derived equations is guaranteed. 
%Here we have locate an other problem of this ansatz, it works only in the non relativistic limit. Again the majority of the gravitoelectromagnetic phenomena are non relativistic and in general we have not a big problem but we would like to have a theory that works if we have a source with relativistic velocities. 

Finally, we can give the equation of motion of a test particle that moves in the field area. First of all the line element has the same form as in the previous section Eq. (\ref{line-element-GEM})
  %%%%%%%%%%%%%%%%%%
\begin{align}
ds^2=c^2\left(1+\frac{2\Phi}{c^2}\right) dt^2-\frac{4}{c}\,(\vec{A} \cdot d\vec{x})\,dt
- \left(1-\frac{2\Phi}{c^2}\right) \delta_{ij} dx^i dx^j.\nonumber
\end{align}
%%%
The Christoffel symbols are also the same as in the previous section: 
%%%%%%%%%%%%%%
\begin{eqnarray}
\G^i_{00} &=& \nonumber
%\frac{1}{2}\,\h^{ij}\left(\frac{2}{c}\, \aa_t h_{0j}-\aa_j h_{00}\right)=
-\frac{1}{c^2}\,\partial^i \Phi
+\frac{2}{c^3}\,\partial_t A^i\,, \\[2mm]\nonumber
%%%%
\G^i_{0j} &=& 
%\frac{1}{2}\,\h^{ik}\left(\an[j]h_{0k}-\an[k]h_{0j}\right)=
\frac{1}{c^2}\,F_{j}^{\2\, i} - 
\,\delta^i_j \,\frac{1}{c^3} \partial_t \Phi, \\[2mm]\nonumber
%%%%
\G^i_{kj} &=& 
%\frac{1}{2}\,\h^{il}\left(h_{lj,k}+h_{lk,j}-h_{jk,l}\right)=
-\frac{1}{c^2}\left(\delta^i_j \,\partial_k \Phi +
\delta^i_k \,\partial_j \Phi -\h_{kj} \,\partial^i \Phi \right).\nonumber
\end{eqnarray}
%%%%%%%%%%%%%%%
In the second of the above equations, we use the definition 
$F_{ij} \equiv \partial_i A_j -\partial_j A_i$.
  Then, the equations on motion, in a familiar vector notation, are \cite{Bakop}:
  %%%%%%%%%%%%%%%
\begin{equation}
m\,\vec{a}=\vec{F}=m\,\vec{E}\left(1+\frac{|\vec{u}|^2}{c^2}\right)
+\frac{4m}{c}\,\vec{u}\times \vec{B} +
2m\left[\frac{\vec{u}}{c}\,\frac{\partial_t \Phi}{c} -
\frac{\vec{u}}{c}\,\left(\frac{\vec{u}}{c} \cdot \vec{E}\right)\right]
\label{Lorentz-GEM}
\end{equation}
%%%%%%%%%%%%%%
 where,  we have set
$\partial_t {\vec A}=0$ and thus ${\vec E}=-\vec{\nabla} \Phi$. The Gravitomagnetic Field ($\vec{B}$) and the gravitoelectromagnetic tensor ($F_{\m\n}$) are related through the equation $F_{ij}=-2\,\e_{ijk}B_k$.
If we are in the non relativistic limit and the scalar potential is static, the above equation reduces to the analogue of the Lorentz force law:
%%%%%%%%%%%%%%%
\begin{equation}
m\,\vec{a}=\vec{F}=m\,\vec{E}+\frac{4m}{c}\,\vec{u}\times \vec{B}\,.
\label{Lorentz-Mashhoon}
\end{equation}
%%%%%%%%%%%%%% 
 
In this section, we have performed again all the calculations of the traditional ansatz without the imposition of a gauge condition in order to investigate in depth the problems that arise.  The nature of the problems forces us to seek modified ansatzes for the metric perturbations. In the following section we   present an alternative ansatz and in the  following chapter we will add corrections to both ansatzes. 
%%%%%
\section{An alternative ansatz}\label{C} 
%%%%%
As we saw in the previous section (\ref{B}), a possible remedy for the problems which appear at the traditional ansatz is the adoption of a new one. In this section we   follow the literature \cite{Bakop}\cite{Costa} and   present an additional ansatz for the metric perturbations. The ansatz we use here has been carefully chosen in order to fix the Lorentz force law. 

The metric perturbations in this section will be:
%%%%%%%%%%%%
\beq 
\tilde h_{00}=\frac{\Phi}{c^2}\,, \qquad \tilde h_{0i}= \frac{A_i}{c^2}\,,
\qquad \tilde h_{ij}=-\frac{\Phi}{c^2}\,\h_{ij}\,, \label{case3}
\eeq
%%%%%%%%%%%
%%%%%%%%%%%%%%%%%%
with the basic difference being the assumption of a non-vanishing (or sub-dominant) $\hh_{ij}$.

Working as before, we first derive the components of $\tilde h_{\mu\nu}$ in mixed
form  
%%%%
\beq 
\tilde h_0^{\2 0}=\frac{\Phi}{c^2}\,, \qquad \tilde h_0^{\2 i}= \frac{A^i}{c^2}\,, \qquad \tilde h_i^{\2 0}= \frac{A_i}{c^2}\,,
\qquad \tilde h_i^{\2j}=-\frac{\Phi}{c^2}\,\delta_i^{\2j}\,,
\eeq
%%%
and then, from the field equations (\ref{field_eqs_full}), we obtain the following
system
%%%%%%%%%%%%%%
\begin{itemize}

\item For $\m=0$ and $\n=0$:

\beq\label{PoissonCase02}
0 = 2 k \rho\,u_0\,u_0\,, 
\eeq

\item  For $\m=0$ and $\n=i$:

\begin{equation}\label{Max2}
\frac{1}{c^2}\,\partial_i\left(\partial_k A^k\right)
-\frac{1}{c^2}\,\delta^{kl}\,\partial_k \partial_l A^i = 2k \rho\,u_0\,u_i\,.
\end{equation}

\item Finally for $\m=i$ and $\n=j$:
\begin{align}\label{ExtraCase02}
-&\frac{1}{c^3}\,\partial_t \left(\partial_i A^j +\partial_j A^i\right)
-\frac{2}{c^2}\,\partial_i\partial_j \Phi  \\[2mm] \nonumber
+&\delta_{ij}\left[\frac{2}{c^2}\,\delta^{kl}\,\partial_k\partial_l \Phi + 
\frac{2}{c^3}\,\partial_t (\partial_k A^k)\right] = 2k \rho\,u_i\,u_j\,.
\end{align}

\end{itemize}
%%%%%%%%%%%%%%%%

Setting again $k \equiv 8\pi G/c^4$, Eq. (\ref{Max2}) takes a form 
similar to that of the fourth Maxwell's equation for both a static scalar
and   vector potential, $\partial_t \Phi=\partial_t \vec{A}=0$,
%%%%%%%%%%%%%%%%
\beq
\vec{\nabla}\,(\vec{\nabla} \cdot \vec{A})
-\nabla^2 \vec{A} = 16\pi G \rho \,\frac{\vec{u}}{c}\,.
\label{4Maxwell02}
\eeq
%%%%%%%%%%%%%%%%
This equation differs from the exact Maxwell equation by a factor of 4 on the
right-hand-side but, as we will see, this will be irrelevant. 
Equation (\ref{PoissonCase02}), that in the previous case gave us Poisson's law,
is now reduced to the trivial result $\rho=0$, which, when combined with the fact
that $\partial_t \vec{A}=0$, leads to the demand that $\Phi$ satisfies the
equation 
%%%%%%%%%%%%%%%%
\beq
\nabla^2\,\Phi=0\,.
\label{Poisson002}
\eeq
%%%%%%%%%%%%%%%
We conclude that this choice for the gravitational perturbations leads to
a static model of gravity in vacuum. For $\rho=0$, the right-hand-side of 
Eq. (\ref{4Maxwell02}) also vanishes making the numerical factor
irrelevant. In retrospect, our assumption of non-vanishing $\tilde h_{ij}$
seems justified: although this component is indeed significantly suppressed
in the presence of an energy-momentum tensor of the form $T_{\mu\nu}=\rho u_\mu u_\nu$,
in vacuum all components are of the same order. 

Imposing the transverse gauge condition ${\hh^{\m\n}}_{\4,\n}=0$ results into
two constraints, namely:

\begin{itemize}

\item For $\m=0$, the Lorentz gauge condition:

\begin{equation}\label{Lorentza1}
\frac{1}{c}\,\an[t]\F + \vec{\nabla}\cdot \vec{A}=0 \4 .
\end{equation}

\item For $\m=i$, the gauge condition dictates that the gravitoelectric field is vanishing:
\begin{equation}\label{zere}
\frac{1}{c}\,\partial_t \vec{A} + \vec{\nabla} \Phi=0\,,
\end{equation}

\end{itemize}
%%%%%%%%%%%%
We note that the different numerical coefficients that
appear in the ansatz, compared to the traditional one (\ref{case1})
used in GEM, allow us to define the GEM fields $\vec{E}$ and $\vec{B}$  in
an exact analogy with the electromagnetism
%%%%%%%%%%%%%%%
\beq
\vec{E} \equiv -\frac{1}{c}\,\partial_t \vec{A} -
\vec{\nabla} \Phi\,, 
\qquad \vec{B} \equiv \vec{\nabla} \times \vec{A}\,.
\eeq
%%%%%%%%%%%%%%%

Although this case seems to be not particularly rich in content compared
to the one studied in the previous  section, it is in contrast free of
additional constraints. The set of equations (\ref{ExtraCase02}) -- both the
diagonal and off-diagonal components -- are trivially satisfied if one uses
the aforementioned constraints Eqs. (\ref{Lorentza1}-\ref{zere}) following from the gauge
conditions. However, we have encountered a big problem in this ansatz, the vanishing of the gravitoelectric field  $\vec{E}$. Taking in consideration that this ansatz is   valid only in   vacuum, we can conclude that   it is very limited. We are only left with the analogue of the magnetostatic field in the Coulomb gauge:   if the gravitoelectric field  $\vec{E}$  vanishes, the scalar potential  $\F$  is a constant; then,  Eq. (\ref{Lorentza1})    reduces  to $\vec{\nabla}\cdot\vec{A}=0$, which is the well known --from the electromagnetism-- Coulomb gauge.

In this ansatz the  spacetime line-element assumes the simplified form 
%%%%%%%%%%%%%%%%%%
\beq
ds^2=c^2\left(1+\frac{2\Phi}{c^2}\right) dt^2-\frac{2}{c}\,(\vec{A} \cdot d\vec{x})\,dt
- \delta_{ij} dx^i dx^j\,.
\label{line-element-case3}
\eeq
%%%%%%%%%%%%%%%%%%

Turning finally to the geodesics equation, by using the expressions for
the initial perturbations $h_{\mu\nu}$ as these are read in the line-element
(\ref{line-element-case3}), we arrive at particularly simple forms
for the Christoffel symbols
%%%%%%%%%%%%
\beq\label{chriscas3}
\G^i_{00} = 
%\frac{1}{2}\,\h^{ij}\left(\frac{2}{c}\, \aa_t h_{0j}-\aa_j h_{00}\right)=
\frac{1}{c^2}\,\partial_i \Phi
+\frac{1}{c^3}\,\partial_t A^i\,, \quad
%%%%
\G^i_{0j}= 
%\frac{1}{2}\,\h^{ik}\left(\an[j]h_{0k}-\an[k]h_{0j}\right)=
\frac{1}{2c^2}\,F_{ij}\,,  \qquad \G^i_{kj} =0\,.
\eeq
%%%%%%%%%%%%
Substituting these into the geodesics equation (\ref{geodesics1-GEM}), we
obtain the exact functional analogue of the Lorentz force  
%%%%%%%%%%%%%%%
\begin{equation}
m\,\vec{a}=\vec{F}=m\,\vec{E}+\frac{m}{c}\,\vec{u}\times \vec{B}\,.
\label{Lorentz-alter2}
\end{equation}
%%%%%%%%%%%%%%
We have seen that the   gravitoelectric field  $\vec{E}$  in this ansatz   vanishes and thus the scalar potential is a constant. Having that in mind, in Eq. (\ref{chriscas3}) the $\G^i_{00}$   should  vanish  and consequently Eq. (\ref{Lorentz-alter2})   should be:
%%%%%%%%%%%%%%%
\begin{equation}
m\,\vec{a}= \,\frac{m}{c}\,\vec{u}\times \vec{B}\,.
\label{Lorentz-alter3}
\end{equation}
%%%%%%%%%%%%%%
In short, we have given the equations of motion in the form of Eq. (\ref{Lorentz-alter2}) in order to show that in this ansatz we can get the correct form of the Lorentz force law. In this case additional terms do not emerge, not even sub-dominant ones in the
non-relativistic limit. We can also see that the same coefficient appears in front of the gravitoelectric and
gravitomagnetic terms, in exact analogy to electromagnetism. Therefore we conclude that
this particular ansatz for the gravitational perturbations may lead to
another class of phenomena observed in vacuum in the context of
gravitoelectromagnetism: the field equations in conjunction to the gauge
condition lead to a self-consistent set of fundamental equations with
no additional constraints and a remarkable similarity to the 
corresponding formulae of electromagnetism. 

However this ansatz has two serious problems; the gravitoelectric field  vanishes and it works only in   vacuum. One could claim that the vanishing of the gravitoelectric field is the most important problem but actually the most important problem is the latter. Usually when we solve a problem in   vacuum, we have information about the source in the form of boundary conditions. In other words, we solve the field equations inside the source and outside (in the vacuum) and we demand that on the boundary (the surface that encloses the source) the two solutions coincide. In the gravitoelectromagnetism theory, we solve the field equations of General Relativity around a rotating massive body, which means that again we have a source. However, in this ansatz we do not have any information about the source; we only know that the field exists, and possibly propagates, in vacuum but we are unable to connect it with its  source.   Despite the problems we have encountered, in this model we have obtained some promising results such as the fact that we  get the exact form of the Lorentz force law. We will continue presenting corrections for the two ansatzes. In fact, as we will see in the following chapter (\ref{E}), the work we presented in this section will lead  us to the best ansatz for the metric perturbation, in which we can have both time-dependent fields and equations of   motion very close to the form of the Lorentz force law.

\clearpage
\thispagestyle{empty}

\chapter{Improving  Gravitoelectromagnetism}\label{4}

\section{A correction to the traditional ansatz for the  metric perturbations}\label{D}

In this section we will try to fix the problems we have encountered in the study of the traditional ansatz. We    stay in the traditional ansatz and assume non-vanishing $\tilde h_{ij}$ components of the metric perturbations.   If we look more carefully at the spatial component of the gauge condition, (Eq. (\ref{add_gauge1})), in section (\ref{A}), we   see that the use of these components of the metric perturbations can instantly solve the problem of the time dependence of the vector potential  $\vec{A}$. In  Mashhoon's works \cite{Mashhoon}\cite{Mash2}\cite{Mash3}\cite{Mashtime}, the spatial components of the gauge condition  ($\hh^{i\m}_{\,\,\,\,,\m}$) are not taken into account and the vector potential is assumed to be time dependent. However in   these articles  the $\hh_{ij}$ components are not zero and, indeed, the vector potential may have time dependence, although nothing is said about the $\hh^{i\m}_{\,\,\,\,,\m}$ condition and its importance on the time dependence of the vector potential. What was  only mentioned here \cite{Mash2} is that these equations involve $\O(c^{-4})$ terms and for this reason they are negligible.

 Now let us think about the form of the $\tilde h_{ij}$ components. According to the usual assumptions of General Relativity, the scalar potential
$\Phi$ is associated only with the $\tilde h_{00}$ component of the metric
perturbations.  To preserve the analogy with  electromagnetism, the vector
potential should appear linearly in the expression of $\tilde h_{\mu\nu}$, and 
thus can only be accommodated by the $\tilde h_{0i}$ component. We could also introduce a scalar potential $\Theta$ in the $\tilde h_{0i}$ components of the metric perturbations through the relation:
\begin{equation}
\hh_{0i}\,=\, \frac{2 \,\Theta}{c^2} A_i. 
\end{equation}
However, as we can easily see   by making the change $A^i\longrightarrow \Theta A^i$ in all equations of the previous chapter involving the $A^i$ potential, that would make the situation worse. As for the  $\tilde h_{ij}$ components, they can  be either zero, as they were in   section (\ref{B}), or    associated with a scalar potential and a tensor potential. The scalar potential  could be a general scalar potential or the gravitoelectric scalar potential $\F$ as we will see in the next section.
Although this may  seem peculiar at first, in the course of our analysis it will
be justified and shown to exhibit interesting features.

We   make the following changes to the traditional ansatz for the metric perturbations $\tilde h_{\mu\nu}$ \cite{Costa}\cite{Natario}\cite{carol}\cite{bako}
%%%%%%%%%%%%
\beq 
\tilde h_{00}=\frac{4\Phi}{c^2}\,, \qquad \tilde h_{0i}=\frac{2A_i}{c^2}\,,
\qquad \tilde h_{ij}=\frac{2\l}{c^4}\,\h_{ij}\,+\,\frac{2}{c^4}\,d_{ij}\,. \label{case4}
\eeq
%%%%%%%%%%%
In the above ansatz the $\hh_{00}$ and $\hh_{0i}$  components of the metric perturbations are again expressed  in terms of $\F$ and $\vec{A}$ potentials, while the  $\tilde h_{ij}$
component consists of a scalar field $\l$ and a traceless symmetric tensor field $d_{ij}$. Obviously the potential $\l$ is connected with the trace of the component $\hh_{ij}$ 
\begin{equation}
\h^{ij}\hh_{ij}=\frac{6}{c^4}\l.
\end{equation}
We can associate the traceless symmetric tensor field $d_{ij}$ with two 	
perpendicular 3-vectors $\vec{C}$ and $\vec{D}$ through the relation 
%%%%
\begin{equation}
d_{ij}\,=\,C_iD_j\,+\,C_jD_i. 
\end{equation}
Then we could easily write our final equations in a familiar vector notation, but for now we  keep the tensor formulation and we   study the most general case.
In some equations, as the equations of motion, the $\tilde h_{ij}$ components   will still be negligible, due to the $1/c^4$ factor, especially in the non relativistic limit, but they will not be  negligible everywhere, as we will shortly see. 
%As we will see these components give us important information and we should use them from the beginning of our work.
%%

From this section we will  start using again the  gauge condition ${\hh^{\m\n}}_{\4,\n}=0$.  The transverse gauge now gives:
%%%%%%%%%%%%%%%%%%%%%%%%%%%%%%%%%%%
\begin{itemize}

\item For $\m=0$, the analogue of the Lorentz gauge condition of the electromagnetism :
\begin{equation}
\frac{1}{c}\,\aa_t \F + \aa_i \left(\frac{A^i}{2}  \right)=0.
\end{equation}\label{lorentz_cond01}

\item And for $\m=i$, an additional condition that connects the scalar $\l$ and the tensor field $d_{ij}$ with the time derivative of the vector potential $\vec{A}$:

\begin{equation}\label{add_gauge01}
\frac{1}{c}\,\aa_j \, \left(\l \h^{ij}+d^{ij} \right) \,=\,-\,\aa_t\,A^i.\,\,
\end{equation}

\end{itemize}
%%%%%%%%%%%%%%%%%%%%%%

 The field equations are:
%%%%%%%
\begin{itemize}

\item For $\m=0 $ and $\n=0\,$:
\begin{equation}\label{poison01}
\da \F\,=\,4 \p G\, \r.
\end{equation}

\item For $\m=0 $ and $\n=i\,$:
\begin{equation}\label{dianysm_01}
\da \left(\frac{A^i}{2} \right)\,=\,\frac{4\p G}{c}\,j^i.
\end{equation}

\item Finally for $\m=i $ and $\n=j\,$:
\begin{equation}\label{add_01}
\frac{1}{2}\,\da \, \left( \l \h^{ij}+d^{ij} \right) \,=\,4\p G\,\,j^i\,u^j\,.
\end{equation}
\end{itemize}
%%%%
Equations (\ref{poison01}) and (\ref{dianysm_01})  have  the same form as the corresponding equations of electromagnetism. Equation (\ref{add_01}) is the Poisson's equation that gives us the additional fields $\l$ and $d_{ij}$. These   are important, despite the fact that they are suppressed due to the $1/c^4$ factor,  because without them Eq. (\ref{add_gauge01}) takes the form $\aa_t\,A^i\,=\,0$. We are going to discuss again the importance of the fields $\l$ and $d^{ij}$ at the end of the section.

In analogy with electromagnetism, we can define the GEM fields $\vec{E}$ and $\vec{B}$
in terms of the GEM potentials \cite{Mashhoon}\cite{Bakop}\cite{Costa}:
%%%%%%%%%%%%%%%
\beq
\vec{E} \equiv -\frac{1}{c}\,\partial_t \left(\frac{\vec{A}}{2}\right) -\vec{\nabla} \Phi\,, 
\qquad \vec{B} \equiv \vec{\nabla} \times \left(\frac{\vec{A}}{2}\right)\,. \label{EB_GEM0}
\eeq
%%%%%%%%%%%%%%%
One may easily see that Eqs. (\ref{poison_1}) and (\ref{dianysm_1}) along with Eqs. (\ref{EB_GEM}) reduce to a set of four Maxwell-like equations for the Gravitoelectromagnetic fields:
%%%%%%%%%%%%%%
\beq
\vec{\nabla} \cdot \vec{E} = 4\pi G \rho\,, \qquad 
\vec{\nabla} \times \vec{B}=\frac{1}{c}\,\partial_t \vec{E} + 
\frac{4\pi G}{c}\,\vec{j}\,, \label{FinalGEM_3.1}
\eeq
and
\begin{equation}\label{FinalGEM_3.2}
\vec{\nabla} \times \vec{E}=\frac{1}{c}\,\partial_t \vec{B}, \4\4 \vec{\nabla} \cdot \vec{B} =0.
\end{equation}
The above equations along with Eq. (\ref{add_01}) describe the gravitational field of a rotating massive body in the weak-field (linear) approximation. 

The spacetime line-element, in this ansatz,
assumes the form 
%%%%%%%%%%%%%%%%%%
\begin{align}\label{line-element-GEM3}
ds^2&=c^2\left(1+\frac{2\Phi}{c^2}-\frac{3\l}{c^4}\right) dt^2-\frac{4}{c}\,(\vec{A} \cdot d\vec{x})\,dt
- \left(1-\frac{2\Phi}{c^2}+\frac{3\l}{c^4} \right) \delta_{ij} dx^i dx^j\nonumber \\[2mm] &+\frac{2}{c^4}d_{ij}dx^i dx^j \,.
\end{align}

Finally, we can find the equation of motion of a test particle
propagating in the background (\ref{line-element-GEM3}). This has  been derived
and discussed in the literature  \cite{Mashhoon}\cite{Bakop}\cite{Costa}\cite{Mash3}, but only in a very simplified
form. In \cite{Mashhoon} the equation of motion is presented only in the non-relativistic limit. In \cite{Bakop} and \cite{Mash3}, although the corrections for large velocities are given,   these do not contain the corrections for the $\hh_{ij}$ components. In \cite{Costa} the authors start with an ansatz, different from the one we are using here, which contains the components $\hh_{ij}$ but at the end of their analysis they assume that  $d_{ij}$ vanish.  Here, we keep all corrections and   comment on their importance at the final stage of our calculation. 

%%%%%%%%%%%%%%%
In the linear approximation, the Christoffel symbols are given by Eq. (\ref{Christoffel}). 
Reading the form of the initial perturbations $h_{\mu\nu}$ from the line-element
(\ref{line-element-GEM3}), we find:
%%%%%%%%%%%%%%
\begin{align}
\G^i_{00} =& 
%\frac{1}{2}\,\h^{ij}\left(\frac{2}{c}\, \aa_t h_{0j}-\aa_j h_{00}\right)=
-\frac{1}{c^2}\,\partial^i \Phi
+\frac{2}{c^3}\,\partial_t A^i\,-\frac{3}{2c^4}\aa^i\l, \\[2mm]
%%%%
\G^i_{0j} =&\4 
%\frac{1}{2}\,\h^{ik}\left(\an[j]h_{0k}-\an[k]h_{0j}\right)=
\frac{1}{c^2}\,F_{j}^{\2\, i} - 
\,\delta^i_j \,\left(\frac{1}{c^3} \partial_t \Phi+\frac{3}{2c^5}\,\aa_t \l\right) +\frac{1}{c^5}\aa_t\, d^i_{\,\,j} \,, \\[2mm]
%%%%
\G^i_{kj} =& 
%\frac{1}{2}\,\h^{il}\left(h_{lj,k}+h_{lk,j}-h_{jk,l}\right)=
-\frac{1}{c^2}\left(\delta^i_j \,\partial_k \Phi +
\delta^i_k \,\partial_j \Phi -\h_{kj} \,\partial^i \Phi \right)\nonumber\\[2mm] 
&-\frac{3}{2c^4}\left(\delta^i_j \,\partial_k \l +
\delta^i_k \,\partial_j \l -\h_{kj} \,\partial^i \l \right)\nonumber\\[2mm]
&+\frac{1}{c^4}\,\left(\aa_k d^i_{\,\,j}\,+\aa_j d^i_{\,\,k}\,-\,\aa^i d_{jk}\,\right).
\end{align}
%%%%%%%%%%%%%%%
%In the second of the above equations, we use the definition $F_{ij} \equiv \partial_i A_j -\partial_j A_i$.

Substituting the above in the geodesics equation  Eq. (\ref{geodesics1-GEM}) we find:
%%%
\begin{align}\label{lorfull}
\ddot{x}^{\,i}=\,&\4E^i +\frac{2}{c}F^{ij}u_j+\frac{1}{c^2}\,\left(\,\frac{5}{2}\aa^i \l+ \aa_j\,d^{ij}+2u^i\aa_t \F +2u^i u^k \aa_k \F\,\right)\nonumber\\[2mm]
&-\frac{1}{c^2}u^k u_k \aa^i \F\, -\frac{1}{c^4} \left(3 u^i \aa_t \l -3u^i u^k \aa_k \l +u^k u_k \aa^i \l \,\right)\nonumber\\[2mm]
&-\frac{1}{c^4}\left(\,2u_j \aa_t d^{ij} +2u^k u^j \aa_k d^i_{\,\,j} - u^j u^k \aa^i d_jk \,\right).
\end{align}
%%%%
The above equation in the non relativistic limit takes the well-known form of the Lorentz force law:
%%%%%
\begin{equation}\label{lorel}
\ddot{x}^{\,i}=\, E^i +\frac{2}{c}F^{ij}u_j,
\end{equation}
%%%%
in which the gravitoelectromagnetic fields  can be time dependent.

In this section we have made a correction to the traditional ansatz for the metric perturbations. We have used the $\tilde h_{ij}$ components, which in previous works were supposed to be negligible due to the $1/c^4$ factor, and we propose that this fixes the time dependence of the potentials. We have also seen that this ansatz keeps the analogy with the true electromagnetism but with the presence of additional fields, i.e  the fields which are produced from the potentials $\l$ and $d_{ij}$. The additional potentials are necessary and we should  have used them from the start of our analysis. Actually the existence of these potentials comes from the fact that Gravity is a Spin 2 field instead of Electromagnetism which is a Spin 1 field. Consequently we   need more degrees of freedom to describe gravity instead of electromagnetism, in which we need initially only four degrees of freedom. When we take a theory of many degrees of freedom, like Gravity, and we use only some of them, we do not use all the information we may get; for this reason we have encountered many problems in the previous chapter. However, this ansatz is still giving us extra terms in the equation of motion. In the next section we will  try to fix the problem of the extra terms in the equation of motion using a modified version of the ansatz we  introduced in section (\ref{C}).

\section{A correction to the alternative ansatz for the metric perturbations}\label{E}

In this section we will  try to make some corrections to the ansatz that we  presented in section (\ref{C}). This section is split into two subsections. In the first one we present the most general case of the ansatz and in the second we give a special case from which we obtain some interesting results.

\subsection{The general case}\label{E1}

In section (\ref{C}) we worked with an ansatz that may give the correct form of the analogue of the Lorentz force law for the equations of motion. However we have encountered many important problems such as the vanishing of both the source and the gravitoelectric field and of the time dependence of the fields. In this section we   try to modify the ansatz of section   (\ref{C}) in order to fix its problems. If we think as in the previous section, the only thing that we have to do is to involve in our analysis the additional degrees of freedom of the gravitational field.  In the ansatz of section (\ref{C}) the components $\hh_{ij}$ are not zero, see Eq. (\ref{case3}); however from  the line element in this case, Eq. (\ref{line-element-case3}), we   see that the $h_{ij}$ components again vanish. But these are the components that we should involve,  in our analysis and associate them with the scalar potential $\l$ and the traceless tensor potential $d_{ij}$.

The ansatz we will use here is the following:
%%%%%%
\beq 
\tilde h_{00}=\frac{\Phi}{c^2}-\frac{3\l}{c^4}\,, \qquad \tilde h_{0i}=\frac{A_i}{c}\,,
\qquad \tilde h_{ij}=-\left(\frac{\F}{c^2}+\frac{\l}{c^4}\right)\,\h_{ij}\,+\,\frac{2}{c^4}\,d_{ij}\,. \label{case5}
\eeq
%%%%%%
In this ansatz the magnitude of the potentials $\F$, $\l$ and $d_{ij}$ is in principle the same. The different $1/c$ coefficients, first,  are connected with the $T_{ij}$ components of the energy-momentum tensor and, second,  will result in simplified forms of  the equations of motion. %The form of Eqs. (\ref{case5})  line element (\ref{line-element-case5}). 
The line-element in this ansatz is:
\begin{align}\label{line-element-case5}
\nonumber ds^2=&\,\,\,\,c^2\left(1+\frac{2\Phi}{c^2}\right) dt^2- 2 \,(\vec{A} \cdot d\vec{x})\,dt \\[2mm] 
&+ \left[\left(1+\frac{2\l}{c^4}\right)\,\h_{ij}+\frac{2}{c^4}\,d_{ij}\,\right] dx^i dx^j\,,
\end{align}
%and it is now profound how we have chosen this ansatz.

We start our analysis from the gauge condition. The transverse gauge condition ${\hh^{\m\n}}_{\4,\n}=0$ gives:

\begin{itemize}

\item For $\m=0$,   an equation similar to the Lorentz gauge condition:
%%%%%%%%
\begin{equation}\label{lor5}
\aa_t\left(\frac{\Phi}{c^3}-\frac{3\l}{c^5}\right)+\aa_i\,\frac{A^i}{c}\,=\,0. 
\end{equation}
%%%%%%%%

\item And for $\m=i$, an equation which relates the gravitoelectric field with the gradient of the scalar $\l$ and divergence of the tensor $d_{ij}$:
%%%%%%%%
\begin{equation}\label{add5}
\aa_t A^i -\aa^i\F\,=\,E^i\,=\,\frac{1}{c^2}\left(\aa^i\l-2\,\aa_jd^{ij}\right).
\end{equation}
%%%%%%%%%
  
\end{itemize}  
From Eq. (\ref{lor5}) we can get the Lorentz gauge condition if we demand that the scalar potential $\l$ is not depending on time $\aa_t\l=0$. Since the Lorentz gauge condition is important for our analysis, from here we   conclude that the scalar potential $\l$ must be static. Although this may seem a problem, considering that we have inserted the scalar $\l$ and the tensor $d_{ij}$ potentials in order to avoid the  problems with the $\F$ and $\vec{A}$ potentials, it is not  a serious one. From Eq. (\ref{add5}) we get an additional constraint for the scalar $\l$ and the tensor $d_{ij}$ potentials; they must be related with the gravitoelectric field. The last equation saves both the gravitoelectric field and the time dependence of the potentials.

We can now continue with the field equations Eq. (\ref{field_eqs}), which inside the source are
\begin{itemize}

\item For $\m=0 $ and $\n=0\,$:
\begin{equation}\label{poison5}
\da \left( \frac{\Phi}{c^2}-\frac{3\l}{c^4}\right)=\,-\frac{16 \p G}{c^2}\,\r.
\end{equation}

\item For $\m=0 $ and $\n=i\,$:
\begin{equation}\label{dianysm5}
\da A^i\,=\,-\frac{16 \p G}{c^2}\,j^i.
\end{equation}

\item Finally for $\m=i $ and $\n=j\,$:
\begin{equation}\label{adeq5}
\da \, \left[-\left(\frac{\F}{c^2}+\frac{\l}{c^4}\right)\,\h_{ij}\,+\,\frac{2}{c^4}\,d_{ij} \right] \,=\,-\frac{16 \p G}{c^4}\,j^i\,u^j\,,
\end{equation}
and if we use Eq. (\ref{lor5}), Eq. (\ref{adeq5}) takes the form:
%%%%%%%
\begin{equation}\label{adeq51}
\da \left( 2\,\l\h^{ij}\,+\,d^{ij} \right)=\,-8\,\p G \left( j^i u^j + \r c^2 \h^{ij} \right).
\end{equation} 

\end{itemize}
%%%%
While, in vacuum the field equations take the form:
\begin{itemize}

\item For $\m=0 $ and $\n=0\,$:
\begin{equation}\label{poison5vac}
\da \Phi\,=\,0.
\end{equation}

\item For $\m=0 $ and $\n=i\,$:
\begin{equation}\label{dianysm5vac}
\da A^i\,=\,0.
\end{equation}

\item Finally for $\m=i $ and $\n=j\,$:
\begin{equation}\label{adeq5vac}
\da \l\,=\,0 \4 and \4 \da d_{ij}\,=\,0.
\end{equation}
\end{itemize}
%%%
The above equations are the field equations which describe the gravitational field around (and inside) a rotating massive body. Here we face the only real problem with this ansatz:   in spite of the fact that we have restored the source,   as we can see from Eqs. (\ref{poison5}-\ref{adeq51}), the field equations inside it have a different form from the true electromagnetic ones. However, in   vacuum the field equations (\ref{poison5vac}-\ref{adeq5vac}) have the correct form. As we  see in the next subsection, we can easily fix this problem.

Finally we can give the equation  of motion of a test particle moving in the background (\ref{line-element-case5}). Here, as always, we   keep all the corrections and we  comment on their importance at the final stage of the calculation. 
The Christoffel symbols for the line-element (\ref{line-element-case5}) are:

\begin{align}\label{chriscas5}
\G^i_{00} &= \frac{1}{c^2}\left( \aa_tA^i -\aa^i\F \right). \\ 
%%%%
\G^i_{0j} &= \frac{1}{2} \h^{ik}\left[\frac{1}{c}\,F_{jk}+\frac{2}{c^5}\, \aa_t \left( \, \l \,\h_{jk}\,+\, d_{jk}\,\right) \right]\,.  \\
%%%%
\nonumber \G^i_{kj} &= \frac{1}{c^4}\left( \d^i_j\,\aa_k\l+ \d^i_k\,\aa_j\l -\h_{jk}\,\aa^i\l\right)\, \\ 
 &\,+ \frac{1}{c^4}\left(\aa_k\,d_j^{\,\,\,i} \,+\,\aa_j\,d_k^{\,\,\,i} \,\,-\, \aa^i\,d_{jk}\, \right).
\end{align}
%%%%%%%%%%%%
Substituting these into the geodesics equation (\ref{geodesics1-GEM}), we
obtain the equations of motion
%%%%%%%%%%%%%%%
\begin{align}\label{lortzfullc5}
\ddot{x}^{\,i}&=\, E^i + F^{ij}u_j \,-\frac{1}{c^4}\,\left( 2\,u^i\,\aa_t\l +2\,u^iu^k\,\aa_k\,\l-u^ku_k\,\aa^i\l \right) \nonumber \\
&\,-\frac{1}{c^4}\,  \left( 2\,u^j\, \aa_t d^i_{\,\,\,j}    +2\,u^iu^k\,\aa_k\,d^i_{\,\,\,j} -\,u^ju^k\,\aa^i\,d_{jk} \right).
\end{align}
In the above equation we define the field $\vec{E}$ as:
\begin{equation}
E^i=-\aa_tA^i -\aa^i\F,
\end{equation}
and the field $\vec{B}$ as
\begin{equation}
F_{ij}=4\,\e_{ijk}B_k.
\end{equation}
We did not define  the fields earlier because the analogy with the electromagnetism is broken in this ansatz. We have defined the above fields in the usual way, but we can not recognize  them as the gravitoelectromagnetic fields, i.e  the analogues of the electromagnetic ones.  In the following section we  will say more about the fields. However, as we can easily see,  equation (\ref{lortzfullc5}) contains additional terms but  even at high velocities, due to the presence of the $1/c^4$ factor, it can take a form  similar  to the Lorentz force law:
\begin{equation}\label{lortzc5}
\ddot{x}^{\,i}=\, E^i + F^{ij}u_j \,.
\end{equation}

In this subsection we presented a modified version of  the ansatz we  used in section (\ref{C}). With the new ansatz,  we have  managed to solve the problems that we have encountered in section (\ref{C}). We have obtained  time-dependent potentials, and we have avoided the vanishing of both the gravitoelectric potential and the source. Also this ansatz is the first which gives us the correct form of the equations of motion even at high velocities. However, the ansatz we have used here has still a serious problem: it does not give  us the correct field equations inside the source,  and as a result the analogy with the true electromagnetism is broken.  In the following subsection we   see how we can solve this problem. 

\subsection{An interesting special case}\label{E2}
The ansatz we have used in the previous subsection fixes all the problems we had encountered in section (\ref{C}) but it still has an important problem. Inside the source, as already mentioned, the analogy with the true electromagnetism is broken. We can easily fix  this problem and   this subsection presents how we can do it. 

If we look more carefully at Eqs. (\ref{poison5}-\ref{adeq51}), we can see that the vanishing of the scalar potential $\l$   may fix the problem. In this subsection we  take the ansatz of the previous one and we   demand that $\l\,=\,0$. This means that the trace of the spatial part of the metric perturbations $h_{ij}$ will be vanishing, i.e. $\h_{ij} h^{ij}\,=\,0$. In the previous subsection, we  discussed  the general case of non-vanishing $\l$. In this subsection we disccuss separately the special case with $\l=0$ as it leads to important results that will also be used  in the following chapter. 

The ansatz we will use here for the metric perturbations is the following
%%%%%%
\beq 
\tilde h_{00}=\frac{\Phi}{c^2}\,, \qquad \tilde h_{0i}=\frac{A_i}{c}\,,
\qquad \tilde h_{ij}=- \frac{\F}{c^2}\,\h_{ij}\,+\,\frac{2}{c^4}\,d_{ij}\,. \label{case6}
\eeq
%%%%%%
As in the previous subsection, the above ansatz   is expressed in terms of the  GEM potentials,    $\Phi(x^\mu)$  and $\vec{A(x^\mu)}$,  plus one additional potential, that comes from the $h_{ij}$ components, the traceless tensor potential $d_{ij}(x^\mu)$.

The transverse gauge condition ${\hh^{\m\n}}_{\4,\n}=0$ gives:

\begin{itemize}

\item For $\m=0$,  an equation similar to the Lorentz gauge condition:
%%%%%%%%
\begin{equation}\label{lor6}
\aa_t \frac{\Phi}{c^3} +\aa_i\,\frac{A^i}{c}\,=\,0. 
\end{equation}
%%%%%%%%

\item And for $\m=i$, an equation which relates the gravitoelectric field with the divergence of the tensor potential $d_{ij}$:
%%%%%%%%
\begin{equation}\label{add6}
\aa_t A^i -\aa^i\F\,=\,E^i\,=\,\frac{2}{c^2} \,\aa_jd^{ij} .
\end{equation}
%%%%%%%%%
  
\end{itemize}  
Eq. (\ref{lor6}) has the well-known form of the Lorentz gauge condition. From Eq. (\ref{add6}) we get an additional constraint for the tensor potential $d_{ij}$: it must be related with the gravitoelectromagnetic field. The last equation, as in the previous subsection, saves both the gravitoelectric field and the time-dependence of the potentials. 

Now we can continue with the field equations Eq. (\ref{field_eqs}), which in the  presence of source have the form 
\begin{itemize}

\item For $\m=0 $ and $\n=0\,$:
\begin{equation}\label{poison6}
\da  \Phi  =\,- 16 \p G \,\r.
\end{equation}

\item For $\m=0 $ and $\n=i\,$:
\begin{equation}\label{dianysm6}
\da A^i\,=\,-\frac{16 \p G}{c}\,j^i.
\end{equation}

\item Finally for $\m=i $ and $\n=j\,$:
\begin{equation}\label{adeq6}
\da \, \left(- \frac{\F}{c^2} \,\h_{ij}\,+\,\frac{2}{c^4}\,d_{ij} \right) \,=\,-\frac{16 \p G}{c^4}\,j^i\,u^j\,,
\end{equation}
and if we use Eq. (\ref{lor6}),   Eq. (\ref{adeq6}) takes the form:
%%%%%%%
\begin{equation}\label{adeq61}
\da  \,d^{ij} =\,-8\,\p G\left( j^i u^j + \r c^2 \h^{ij} \right).
\end{equation} 

\end{itemize}
%%%%
Whereas in  vacuum the field equations are
\begin{itemize}

\item For $\m=0 $ and $\n=0\,$:
\begin{equation}\label{poison6vac}
\da \Phi\,=\,0.
\end{equation}

\item For $\m=0 $ and $\n=i\,$:
\begin{equation}\label{dianysm6vac}
\da A^i\,=\,0.
\end{equation}

\item Finally for $\m=i $ and $\n=j\,$:
\begin{equation}\label{adeq6vac}
 \da d_{ij}\,=\,0.
\end{equation}
\end{itemize}
The above equations are the field equations which describe the gravitational field around (and inside) a rotating massive body. As in the previous subsection we have restored the source but, as we can see from Eqs. (\ref{poison6}-\ref{dianysm6}), the coefficient on  the right-hand side  is $16 \pi$, instead of $4 \pi$ which is the coefficient appearing in the true electromagnetism. Despite that, in  vacuum the field equations (\ref{poison6vac}-\ref{adeq6vac}) have the correct form.  We can fix the problem of the wrong coefficient easily by redefining the gravitoelectromagnetic potentials as:
%%%%
\begin{align}\label{redpot}
\F& \longrightarrow 4 \F^*, \nonumber \\
\vec{A}& \longrightarrow 4 \vec{A^*},
\end{align}
%%%
where $\F^*$ and $\vec{A^*}$ are the true potentials. If we employ the new potentials, Eq. (\ref{poison6}) takes the form:
%%%
\begin{equation}\label{pois6}
\da  \Phi^* =\,- 4 \p G \,\r,
\end{equation}
while Eq. (\ref{dianysm6}) takes the form:
\begin{equation}\label{diany6}
\da A^{*i}\,=\,-\frac{4 \p G}{c}\,j^i.
\end{equation}
The above equations have the correct form and along with Eq. (\ref{adeq61}) are the full system of the field equations. 

Now we can define a gravitomagnetic and a gravitoelectric field as:
%%%%%%%%%%%%%%%
\beq\label{ebdef}
\vec{E} \equiv - \,\partial_t \vec{A^*} -
\vec{\nabla} \Phi ^* \,, 
\qquad \vec{B} \equiv \vec{\nabla} \times \vec{A^*}\,.
\eeq
%%%%%%%%%%%%%%%
One may easily see that Eqs. (\ref{pois6}) and (\ref{diany6}) along with Eqs. (\ref{ebdef}) reduce to a set of four Maxwell-like equations for the Gravitoelectromagnetic fields:
%%%%%%%%%%%%%%
\beq
\vec{\nabla} \cdot \vec{E} = 4\pi G \rho\,, \qquad 
\vec{\nabla} \times \vec{B}= \,\partial_t \vec{E} + 
\frac{4\pi G}{c}\,\vec{j}\,, \label{FinalGEM61}
\eeq
and
\begin{equation}\label{FinalGEM62}
\vec{\nabla} \times \vec{E}= \,\partial_t \vec{B}, \4\4 \vec{\nabla} \cdot \vec{B} =0.
\end{equation}
Consequently, this ansatz gives us the correct form of the field equations.

The line-element in this ansatz is:
\begin{align}\label{line-element-case6}
  ds^2= \,c^2\left(1+\frac{2\Phi}{c^2}\right) dt^2- 2 \,(\vec{A} \cdot d\vec{x})\,dt  
 + \left(\h_{ij}+\frac{2}{c^4}\,d_{ij}\,\right) dx^i dx^j\,.
\end{align}

Finally we can give the equations of motion of a test particle moving in the background (\ref{line-element-case6}).  
The Christoffel symbols for the line element (\ref{line-element-case6}) are:
%%%%
\begin{align}\label{chriscas6}
\G^i_{00} &= \frac{1}{c^2}\left( \aa_tA^i -\aa^i\F \right). \\[2mm] 
%%%%
\G^i_{0j} &= \frac{1}{2} \h^{ik}\left(\frac{1}{c}\,F_{jk}+\frac{2}{c^5}\, \aa_t \, d_{jk}\,\right) \,.  \\[2mm]
%%%%
\G^i_{kj} &= \frac{1}{c^4}\left(\aa_k\,d_j^{\,\,\,i} \,+\,\aa_j\,d_k^{\,\,\,i} \,\,-\, \aa^i\,d_{jk}\, \right).
\end{align}
%%%%%%%%%%%%
Substituting these into the geodesics equation (\ref{geodesics1-GEM}), we
obtain the equations of motion
%%%%%%%%%%%%%%%
\begin{align}\label{lortzfullc6}
\ddot{x}^{\,i} =\, 4 \, E^i + F^{ij}u_j \, \,-\frac{1}{c^4}\,  \left( 2\,u^j\, \aa_t d^i_{\,\,\,j}    +2\,u^iu^k\,\aa_k\,d^i_{\,\,\,j} -\,u^ju^k\,\aa^i\,d_{jk} \right).
\end{align}
As we can see, the above equation contains some corrections but, due to the presence of the $1/c^4$ factor even at high velocities, it can take a form similar to the Lorentz force law:
\begin{equation}\label{lortzc6}
\ddot{x}^{\,i}=\, 4\, E^i + F^{ij}u_j \,,
\end{equation}
or in a more familiar vector notation:
\begin{equation}\label{lortzvec6}
\ddot{\vec{x}}\,=\,4\,\left(\vec{E}+\vec{u}\times \vec{B} \right).
\end{equation}
As we can see from the above equation, in this ansatz, the equations of motion are very close to the Lorentz force law. The only difference is an overall factor 4 on the right hand side of the equation. 

In this subsection we have solved the problems that we have  encountered in the previous one by --simply-- setting  $\l=0$. Although this may be a  special case, it has led to the correct form of both the field equations and the Lorentz force law. Moreover, this ansatz does not generate any corrections to the Lorentz force, even at high velocities. Concluding, the ansatz that we have used here is the optimum as it preserves  the analogy between gravitation and electromagnetism.

\chapter{True Electromagnetism}\label{5}

\section{An alternative expression for the Lorentz force law}\label{F}

The usual procedure when someone solves an electromagnetic problem, in the Lorentz gauge, is first to find the potential through the relation
%%%
\begin{equation}\label{max12}
\da A^\m \,=\, \frac{4 \p C}{c^2} \, J^\m ,
\end{equation}
%%%%
where the electromagnetic 4-potential is defined as $A^\m\,=\,\left(\F/c\,,\vec{A}\,\right) $ and $C$  is the Coulomb's constant ($C\equiv\frac{1}{4 \pi \e_0}$). Then one is able to calculate the 
electric and  magnetic fields
\begin{equation}\label{def51}
\vec{E}\,=\,-\vec{\nabla}\F\,-\aa_t\vec{A} \quad\,\, and \quad\,\, \vec{B}\,=\,\vec{\nabla}\,\times \, \vec{A},
\end{equation}
%%%%
and finally to find the equation of motion of a test particle 
%%%%%
\begin{equation}\label{lor51}
m \ddot{\vec{x}}\,=\,q\,\left(\,\vec{E}\,+\,\vec{u}\,\times \, \vec{B} \,\right).
\end{equation}

However, in the previous chapters, we have seen that the Christoffel symbols can give us the definitions of both the Electric and the Magnetic field and that the geodesics equation contains the expression for the Lorentz force law. Consequently, instead of Eq. (\ref{lor51}), we could use a modified version of the  geodesics equation in order to describe the equations of motion.

First of all we use the potential from Eq. (\ref{max12}) in order to define an approximately flat ``space-time" $g_{\m\n}\approx \h_{\m\n}+h_{\m\n}$ with line element
%%%%%
\begin{equation}\label{line51}
ds^2\,=\,\left(1+\frac{2\F}{c^2}\right)c^2dt^2-2\left(\vec{A} \cdot\vec{dx} \right) dt + \h_{ij}dx^i dx^j.
\end{equation}
%%%%%%
Next we claim that the equation 
%%%%%%%%%%%%%%%%%%%
\beq
m\, \frac{d^2 x^\rho}{ds^2} + q\, \Gamma^\rho_{\mu\nu}\,\frac{dx^\mu}{ds}
\frac{dx^\nu}{ds}=0, \label{geodesics-GEM51}
\eeq
%%%%%%%%%%%%%%%%%%%
describes the motion of a test particle with charge $q$ and mass $m$. The above equation is not anymore a geodesics equation, it  gives geodesic curves only if $q=m$. Eq. (\ref{geodesics-GEM51}) takes the explicit form
%%%%%%%%%%%%%%%%
\begin{equation}
\frac{m}{q} \, \frac{d^2x^i}{dt^2}+ c^2\,\G^i_{00}+
2c \,\G^i_{0j}\,\frac{dx^j}{dt}+
\G^i_{kj}\,\frac{dx^k}{dt}\frac{dx^j}{dt} =0\,. \label{geodesics1-GEM51}
\end{equation}
%%%%%%%%%%%%%%% 
The Christoffel symbols Eq. (\ref{Christoffel}) for the above line element  are
%%%%%%%
\begin{align} 
\G^i_{00} &= - \frac{E^i}{c^2}. \label{chriscas511}\\[2mm] 
%%%%
\G^i_{0j} &= \frac{1}{2c} \e_{ijk} B_k \,, \label{chriscas512} \\[2mm]
%%%%
\G^i_{0j} &=0, \label{chriscas513}
\end{align}
%%%%%%%%%%%%
where we have used the definitions for the fields  (\ref{def51}). Finally if we use Eqs. (\ref{chriscas511}-\ref{chriscas513}) we can see that the modified geodesics equation  (\ref{geodesics-GEM51}) is equivalent to the Lorentz force law  (\ref{lor51}).

In this chapter we suggest that we can use the potential of the electromagnetism in order to define a ``geometry", and then from a modified geodesics equation we can get the Lorentz force law. This is just an additional way for someone to find the equations of motion. However, if we use Eq. (\ref{geodesics-GEM51}), we do not actually need to define the electromagnetic fields. We can start from Eq. (\ref{max12}), then we use the potential to define the line element Eq. (\ref{line51}) --bypassing the definition of the fields-- and finally we can find the equation of motion using Eq. (\ref{geodesics-GEM51}). With this new definition of the equations of motion we have connected electromagnetism with a ``geometry". Nevertheless, the potential is still calculated in the usual way  --the well-known equations of electromagnetism-- and we can only construct the geometry (Eq. (\ref{line51})) by hand. In the following sections we will  try to find an alternative formulation for electrodynamics in which the metric tensor   is the solution of the field  equations. In this point we  must say that our  objective is not to replace Maxwell's theory of electromagnetism by a tensorial  theory but rather to investigate how far the analogy between the two dynamics can go.

\section{An alternative description of electromagnetism}\label{G}

In the previous section we have presented   an alternative expression for the Lorentz force law which premises the construction of a ``geometry" using the electromagnetic potential. In this section we   try to find a new formulation for electrodynamics in which the metric tensor is a fundamental quantity.

A careful inspection of Eq. (\ref{line51}) shows that the electromagnetic potential is related with the metric perturbations as follows:
%%%%%%
\begin{equation}
h_{00}=\frac{2\F}{c^2}, \4 h_{0i}=\frac{A_i}{c} \4 and \4 h_{ij}=0.
\end{equation}
%%%%%%%
The above, along with the analysis in  Chapters \ref{3} and \ref{4}, dictates that we should use the formulation of General Relativity in the weak field approximation. We   use, as field equations, Eq. (\ref{field_eqs_full}) 
%%%%%%%%%%%%%%%%
\begin{equation}
{\hh^{\a}}_{\2\m,\n\a}+{\hh^{\a}}_{\2\n,\m\a}-
\da \hh_{\m\n}-\h_{\m\n}\,{\hh^{\a\b}}_{\4,\a\b}=2k\,T_{\mu\nu}\,, \nonumber
\end{equation}
%%%%
in which the proportionality constant $k$ is going to be defined later and the energy-momentum tensor has the form $T_{\m\n}\,=\,\r\,u_\m u_\n$, with $\r$ the electric charge density. 

However if we try to describe electromagnetism with a ``gravity-like" theory a big problem arises. Electromagnetism is a theory with 4 degrees of freedom while gravity has 10. We get a theory very rich in content and we suppress it. 

In a previous work of ours \cite{Bakop} we have used the following ansatz for the metric perturbations $\tilde h_{\mu\nu}$
%%%%%%%%%%%%
\beq 
\tilde h_{00}=\frac{\alpha \hat\Phi}{c^2}\,, \qquad 
\tilde h_{0i}= \frac{\beta \hat A_i}{c^2}\,,
\qquad \tilde h_{ij}=\frac{\gamma \hat \Phi}{c^2}\,\delta_{ij}\,, \label{h-EM}
\eeq
%%%%%%%%%%%
where $\alpha$, $\beta$ and $\gamma$ are arbitrary numerical coefficients while $\hat{\F}$ and $\hat{\vec{A}}$ are the electromagnetic potentials and, as we can easily see, the potentials have the same magnitude. In the previous chapters we have showed that the form of the  metric perturbation affects the derived equations for the scalar and  vector potential. In order to  study simultaneously a large collection of choices we have inserted the  $\alpha$, $\beta$ and $\gamma$  coefficients. We can see that the ansatz of  section (\ref{B}) corresponds to the choice $(\alpha=\beta=1,\gamma=0)$ while the ansatz of  section (\ref{C}) corresponds to $(\alpha=\beta=\gamma=1)$. Our goal is to reduce the field equations of General Relativity to Maxwell's equations. The only way to do this is to demand that the choice of $\hh_{\m\n}$ has to be linear in the electromagnetic potentials  and should not contain any derivatives. Also, we have mentioned that we need only four degrees of freedom, as a result, the additional potentials we have used in Chapter \ref{4}  should apparently vanish  here. We thus conclude  that the scalar $\F$ and the vector $\vec{A}$ potentials can be accommodated in the metric perturbations $\hh_{\m\n}$ in a limited number of distinct ways. With the insertion of the coefficients $\alpha$, $\beta$ and $\gamma$ in the above ansatz Eq. (\ref{h-EM}) we have included all  possible choices.

As in the previous chapters we start our analysis from the linearised form of Einstein's field equations 
(\ref{field_eqs_full}). The field equations for the  ansatz (\ref{h-EM}) take the form
%%%%%%%%%%%%%%
\begin{itemize}
\item For $\m=0$ and $\n=0$
\begin{equation}
\frac{(\alpha-\gamma)}{c^2}\,\delta^{ij}\,\partial_i \partial_j \hat \Phi =
2 k \rho\,u_0\,u_0\,, \label{PoissonCase4}
\end{equation}
\item For $\m=0$ and $\n=i$
\begin{equation}
\frac{(\alpha-\gamma)}{c^3}\,\partial_i \partial_t \hat \Phi +
\frac{\beta}{c^2}\,\partial_i\left(\partial_k \hat A^k\right)
-\frac{\beta}{c^2}\,\delta^{kl}\,\partial_k \partial_l \hat A^i = 2k \rho\,u_0\,u_i\,, 
\label{4Maxwellcase4}
\end{equation}
\item And finally for $\m=i$ and $\n=j$
\begin{align}
\, & -\frac{\beta}{c^3}\,\partial_t \left(\partial_i \hat A^j +
\partial_j \hat A^i\right)
-\frac{2\gamma}{c^2}\,\partial_i\partial_j \hat \Phi \, \nonumber  \\[2mm]
\, & + \,\delta_{ij}\left[\frac{2\gamma}{c^2}\,\delta^{kl}\,\partial_k\partial_l \hat\Phi + 
\frac{2\beta}{c^3}\,\partial_t (\partial_k \hat A^k)+
\frac{(\alpha-\gamma)}{c^4}\,\partial^2_t \hat \Phi\right]
= 2k \rho\,u_i\,u_j\,.
\label{ExtraCase4}
\end{align}
%%%%%%%%%%%%%%%%
\end{itemize}
%%%%%%%%%%%%%%%%

From the above we can see that for $\alpha=\gamma$,  Eq. (\ref{PoissonCase4}) reduces to $\rho=0$ and we only get a model of electro-magnetism in vacuum, a model similar to the one that we have found in section (\ref{C}) in the context of Gravitoelectromagnetism. On the other hand, 
for $\alpha \neq \gamma$,  if we identify the proportionality constant as:
%%%%%%%%%%%%%%%%
\beq
k \equiv - \frac{2\pi C}{c^4}\,(\alpha-\gamma)\,, \label{k-EM}
\eeq
%%%%%%%%%%%%%%
we obtain the Poisson's equation for the scalar potential $\hat \F$
%%%%%%%%%%%%%%%%
\beq
\nabla^2\,\hat\Phi=-4\pi C \rho\,.
\label{Poisson}
\eeq
%%%%%%%%%%%%%%% 
For the above value for $k$, and using the more familiar vector notation,
Eq. (\ref{4Maxwellcase4}) takes the form 
%%%%%%%%%%%%%%%%
\beq
\vec{\nabla}\,(\vec{\nabla} \cdot \hat{\vec{A}} + 
\frac{1}{c}\,\partial_t \hat \Phi)
-\nabla^2 \hat{\vec{A}} = 4\pi C \rho \,\frac{\vec{u}}{c}\,,
\label{4Maxwellstatic}
\eeq
where we have assumed that $\beta=\alpha-\gamma$. Consequently, there is an infinite number of choices  one could make for the
numerical coefficients appearing in the metric perturbations $\tilde h_{\mu\nu}$,
which could derive the electromagnetic equations from the gravitational field
equations (\ref{field_eqs_full}). Now we can define the   electric
$\hat{\vec{E}}$ and magnetic field $\hat{\vec{B}}$, in terms of the scalar $\hat\F$  and the vector $\hat{\vec{A}}$ potentials, in the usual way
%%%%%%%%%%%%%%%
\beq
\hat{\vec{E}} \equiv -\frac{1}{c}\,\partial_t \hat{\vec{A}} -
\vec{\nabla} \hat\Phi\,, 
\qquad \hat{\vec{B}} \equiv \vec{\nabla} \times \hat{\vec{A}}\,. \label{EB_defs}
\eeq
%%%%%%%%%%%%%%%
The above definitions are equivalent to the following Maxwell's equations
%%%%%%%%
\begin{equation}
\vec{\nabla} \cdot \hat{\vec{B}} = 0\,, \qquad 
\vec{\nabla} \times \hat{\vec{E}}=-\,\partial_t \hat{\vec{B}}  \,,\label{otherMaxwells}
\end{equation}
while, it is easy to show  that Eqs. (\ref{Poisson}) and (\ref{4Maxwellstatic}) are
the remaining Maxwell's equations, 
%%%%%%%%%%%%%%
\beq
\vec{\nabla} \cdot \hat{\vec{E}} = 4\pi C \rho\,, \qquad 
\vec{\nabla} \times \hat{\vec{B}}=\frac{1}{c}\,\partial_t \hat{\vec{E}} + 
\frac{4\pi C}{c}\,\vec{j}\,,\label{max554}
\eeq
%%%%%%%%%%%%%
under the constraint that the vector potential is not depending on time,
$\partial_t \hat{\vec{A}}=0$. One can easily see that due to the coefficient $(\alpha-\gamma)$
 in Eqs. (\ref{PoissonCase4}) and (\ref{4Maxwellcase4}),
Maxwell's equations  do not change under the simultaneous changes
$(\alpha \leftrightarrow \gamma)$ and $\hat \Phi \rightarrow -\hat \Phi$. 
Therefore, instead of what happens in Gravitoelectromagnetism, where $\tilde h_{00}$ must always be connected with the Newtonian potential, and as a result it should always be non-vanishing,
in the case that we study here, even if the  $\tilde h_{00}$ component of the metric perturbations is zero and the potential $\hat \Phi$
is introduced only through the $\tilde h_{ij}$ component, we could also  recover the whole set of Maxwell's equations. 
Nevertheless, as in Gravitoelectromagnetism, along with the  Maxwell's equations, the field equations give an additional
set of equations, Eqs. (\ref{ExtraCase4}). For the above choice  of $k$ and $\beta$, these equations take the form
%%%%%%%%%%%%%%%
\bea
\hspace*{-2cm}\frac{1}{c}\,\partial_t \left(\partial_i \hat A^j +\partial_j \hat A^i\right)
+\frac{2\gamma}{\alpha-\gamma}\,\partial_i\partial_j \hat \Phi \nonumber \hspace*{5cm} &&\\[2mm]
\hspace*{1cm}
- \,\delta_{ij}\left[\frac{2\gamma}{\alpha-\gamma}\,\delta^{kl}\,\partial_k\partial_l \hat \Phi + 
\frac{2}{c}\,\partial_t (\partial_k \hat A^k)+
\frac{1}{c^2}\,\partial^2_t \hat \Phi\right]
= 4\pi C \rho\,u_i\,u_j/c^2\,.&& \label{additional4}
\eea
%%%%%%%%%%%%%%%
For $i=j$, the above reduces to the relation
%%%%%%%%%%%%%%
\beq
\partial_t^2 \hat \Phi =- \frac{4\pi C}{3} \rho\,|\vec{u}|^2\,,
\label{diagonalcase1}
\eeq
%%%%%%%%%%%%%%
with $(u^1)^2=(u^2)^2=(u^3)^2$, for any value of $\alpha$ and $\gamma$.
For $i \neq j$, the term proportional to $\delta_{ij}$ in Eq. (\ref{additional4})
  vanishes and,  under the assumption that $\alpha=2\gamma$,  the remaining terms  can be combined to form the components of the electric field. In that case,  Eq. (\ref{additional4}) takes the form
%%%%%%%%%%%%
\beq
\partial_i \hat E_j+\partial_j \hat E_i=- 4\pi C \rho\,u_i\,u_j/c^2\,.
\label{off-diagonalcase4}
\eeq
%%%%%%%%%%%%%%%
In the case where  $\alpha \neq 2\gamma$, Eq. (\ref{additional4}) is formed of
  time and space-derivatives of the electromagnetic potentials $\hat{\vec{A}}$ and
$\hat \Phi$. 

Therefore, due to the tensorial structure of the formalism, the four Maxwell's equations are accompanied by additional constraints. Equivalent constraints have also arisen in the analysis of the previous chapters. In chapter \ref{4} we have showed that, if we use the additional degrees of freedom of the gravitational field, we can solve all the problems  related to the time-dependence of the fields. Here, we use the formalism of General Relativity in order to describe the Electromagnetism and, as we have already mentioned, we can not use more than four of the available degrees of freedom. Consequently, in this case we can not avoid the presence of these constraints and, thus, with this formalism we can only describe a special part of Electromagnetism and not the general case. We can only describe magnetostatics since the vector potential $\vec{A}$  does not depend on time and the distribution of charges should be isotropic as in  section (\ref{B}).

If we impose a gauge condition, we  get additional constraints to the model; some
of these constraints  complete the theory, while others give
unnecessary restrictions to the electromagnetic potentials. As in the previous chapters, we will use the transverse gauge condition $\tilde h^{\mu\nu}_{\4,\n}=0$. The
time-component of the gauge condition takes the  form
%%%%%%%%%%%%%%%%%%
\begin{equation}
\frac{\alpha}{c}\,\partial_t \hat \Phi+ \beta\,\partial_i \hat A^i=0\,.
\end{equation}
%%%%%%%%%%%%%%%%%
If we demand $\alpha=\beta$, the above reduces  to the well known
Lorentz condition 
%%%%%%%%%%%%%%%
\begin{equation}\label{Lor_cond}
\frac{1}{c}\,\partial_t\hat \Phi + \vec{\nabla}\cdot \hat{\vec{A}}=0\, \4 or \4 A^\m_{\,\,\,,\,\m}=0,
\end{equation}
%%%%%%%%%%%%%%%%
where we have defined the electromagnetic 4-potential as $A^\m=\left(\,\F\,,\,\vec{A}\,\right)$. The spatial components of the gauge condition, though, lead to an additional constraint
%%%%%%%%%%%%%%%
\begin{equation}\label{gauge_add}
\frac{\beta}{c}\,\partial_t \hat{\vec{A}} + (\alpha-\beta)\,\vec{\nabla} \hat \Phi=0\,,
\end{equation}
%%%%%%%%%%%%%%%
where we have used the relation $\beta=\alpha-\gamma$. The choice
$\alpha=\beta$ leads to the time-independence of the vector potential $\hat{\vec{A}}$;
in this case we do not get a new constraint, because we already know --from the field equations-- that the vector potential should be static. If we choose
$\alpha=2\beta$, we find that the electric field   
$\hat{\vec{E}}=-\partial_t \hat{\vec{A}}/c -\vec{\nabla} \hat \Phi$
should vanish, a result similar to the one of  section (\ref{C}). For all the remaining cases, Eq. (\ref{gauge_add})
imposes an additional constraint between the scalar and vector potentials
which does not exist in the traditional electromagnetism.  

%%%%%%%edo tha mporouse na mpei kai i parakato paragrafos%%%%%%

%%%%%%%The choice of the gauge condition is of course not unique, therefore theabove analysis is by no means exhaustive; it merely acts in an indicative way regarding the type of the constraints that one would end up with. Another usual gauge condition, the {\it transverse-traceless} one is further supplemented by the demand that the trace of $\tilde{h}$ should be zero which eliminates altogether the scalar potential $\hat \Phi$ from the theory.%%%%%%

Now we can find the equations of motion of a test particle which moves in the electromagnetic field. As in the previous section, we  assume that the equations of motion are given from the modified geodesics equation (\ref{geodesics-GEM51}): 
%%%%%%%%%%%%%%%%%%%
\beq
m\,\frac{d^2 x^\rho}{ds^2} + q\,\Gamma^\rho_{\mu\nu}\,\frac{dx^\mu}{ds}
\frac{dx^\nu}{ds}=0\,. \nonumber
\eeq
%%%%%%%%%%%%%%%%%%%
In the linear
approximation, the  components of the Christoffel symbols that
appear in Eqs. (\ref{geodesics-GEM51}-\ref{geodesics1-GEM51}) have the form
%%%%%%%%%%%%%%
\bea
\G^i_{00} &=& 
%\frac{1}{2}\,\h^{ij}\left(\frac{2}{c}\, \aa_t h_{0j}-\aa_j h_{00}\right)=
\frac{(\alpha+3\gamma)}{4c^2}\,\partial_i \hat \Phi
+\frac{\beta}{c^3}\,\partial_t \hat A^i\,, \\[2mm]
%%%%
\G^i_{0j} &=& 
%\frac{1}{2}\,\h^{ik}\left(\an[j]h_{0k}-\an[k]h_{0j}\right)=
\frac{\beta}{2c^2}\,\hat F_{ij} - 
\frac{(\alpha-\gamma)}{4c^3}\,\delta^i_j \,\partial_t \hat \Phi\,, \\[2mm]
%%%%
\G^i_{kj} &=& 
%\frac{1}{2}\,\h^{il}\left(h_{lj,k}+h_{lk,j}-h_{jk,l}\right)=
-\frac{(\alpha-\gamma)}{4c^2}\left(\delta^i_j \,\partial_k \hat \Phi +
\delta^i_k \,\partial_j \hat \Phi -\delta_{kj} \,\partial_i \hat \Phi \right),
\eea
%%%%%%%%%%%%%%%
where we have defined the spatial part of the electromagnetic tensor as 
$\hat F_{ij}=\partial_i \hat A_j -\partial_j \hat A_i$. The electromagnetic tensor is connected with the magnetic field through the relation $\hat F_{ij}=-\,\e_{ijk}\,\hat{B_k}$.
Substituting the above Christoffel symbols into Eq. (\ref{geodesics1-GEM51}), we find
%%%%%%%%%%%%%%%%%%%
\begin{align}\label{Lorentz-gen}
\, & m\,a^i+q \left[\frac{\beta}{c}\,\partial_t \hat A^i + \frac{(\alpha+3 \gamma)}{4}\,
\partial_i \hat\Phi\right] - \frac{\beta q}{c}\,\epsilon_{ijk}\,u^j \hat B_k   - \frac{(\alpha-\gamma)q}{2 c^2}\,(\partial_t \hat \Phi)\,u^i  \nonumber \\[2mm]  \, &   +
\frac{(\alpha-\gamma) q}{4c^2}\left[\partial_i \hat \Phi\,(u^k u^k) -
2(\partial_j \hat \Phi\,u^j)\,u^i\right]=0\,.
\end{align}
%%%%%%%%%%%%%%%%%%
Under the assumption that we can construct a theory of  electromagnetism using the formalism of General Relativity, the above equation describes the motion of a massive, charged test particle. The above equation is a generalized form of the Lorentz force of the system. This general form   is obviously far away from what we would expect from a realistic electromagnetic theory.

Now let us examine some  special cases. If $\alpha=\gamma$, we already know that we get a static theory in  vacuum. In this case, the last two terms on the left hand side of Eq. (\ref{Lorentz-gen}) and all the time derivatives  trivially vanish and we get the more familiar form
%%%%%%%%%%%%%
\begin{equation}
m\,\vec{a}=\vec{F}=q\alpha\,\hat{\vec{E}}+\frac{q\beta}{c}\,\vec{u}\times \hat{\vec{B}}\,.
\label{Lorentz}
\end{equation}
%%%%%%%%%%%%%
From the above equation we see that  we must choose $\alpha=\beta$, because we know that we need a  common numerical factor in front of the electric and magnetic terms in the Lorentz force. Actually, we should choose $\alpha=\beta=1$   in order to get the correct expression for the Lorentz force, but for any other value of the coefficients $\alpha,  \,\beta $  we could easily fix the problem by  modifying  the 
equation of motion Eq. (\ref{geodesics-GEM51}).

On the other hand, if $\alpha \neq \gamma$, we have seen that we get a more realistic theory of electromagnetism in terms of fields, but in the expression of the Lorentz force  extra terms appear. If we apply the constraint $\beta=\alpha-\gamma$, which we know  is necessary to restore the form of Maxwell's equations, Eq. (\ref{Lorentz-gen}) takes the form
%%%%%%%%%%%%%%%
\begin{equation}
m\,\vec{a}=\vec{F}=\frac{q}{4}\,\hat{\vec{E}}\left[(4\alpha-3\beta)+
\frac{\beta\,|\vec{u}|^2}{c^2}\right]
+\frac{\beta q}{c}\,\vec{u}\times \hat{\vec{B}} +
\frac{\beta q}{2}\left[\frac{\vec{u}}{c}\,\frac{\partial_t \hat\Phi}{c} -
\frac{\vec{u}}{c}\,\left(\frac{\vec{u}}{c} \cdot \hat{\vec{E}}\right)\right].
\label{Lorentz-case4}
\end{equation}
%%%%%%%%%%%%%%
The additional terms which appear  in the above expression are similar to that arising in   section (\ref{B}) in the context of Gravitoelectromagnetism. In the non relativistic limit and if we also assume that the potentials are static, the above equation takes the form
%%%%%
\begin{equation}
\vec{F}=\frac{q}{4}\,(4\alpha-3\beta)\,\hat{\vec{E}}
+\frac{\beta q}{c}\,\vec{u}\times \hat{\vec{B}},
\label{Lorentz-case40}
\end{equation}
%%%%%%%%%%%%%%
which for $\alpha=7\beta/4$ takes the simple form 
%%%%%
\begin{equation}
\vec{F}=\beta q\,\hat{\vec{E}}
+\frac{\beta q}{c}\,\vec{u}\times \hat{\vec{B}}.
\label{Lorentz-case400}
\end{equation}
%%%%%%%%%%%%%%
The above equation for $\beta=1$, takes the correct form of the Lorentz force, but it can take the correct form, even for a general value of $\beta$, if we redefine the equation of motion Eq. (\ref{geodesics-GEM51}). However, this model can not be realistic since it contains corrections for high velocities which we know that  do not exist in the theory of electromagnetism. 

In this section we have showed that  the formalism of the General Relativity is able to describe electromagnetism in some special cases. This happens because the tensorial structure of gravity adds additional unnecessary constraints in the theory or/and extra terms in the Lorentz force law. In the following section we will try to fix these problems   by using the additional degrees of freedom that  the theory contains. 

\section{Employing the additional degrees of freedom}\label{H}

In this section we   use the additional degrees of freedom of the theory (i.e. the $h_{ij}$ components of the metric perturbations) in order to examine if we can get a more realistic theory of electromagnetism. We expect that the use of the additional degrees of freedom   will lead to the appearance of new fields. Therefore, the theory we  present here is a generalized theory of electromagnetism which unifies the classical electromagnetism with the new fields which we expect they arise. If we want this theory to be realistic, the new fields should not have any contribution into observable quantities or, if they contribute, their contribution should be negligible.

In  subsection (\ref{E2}) we showed that the ansatz which we  used   gives us the best results  in the context of Gravitoelectromagnetism. Here we   use the results of the subsection (\ref{E2}) in order to describe the generalized theory of electromagnetism. We use  again  as field equation the linearized equations of General Relativity, Eq. (\ref{field_eqs}),  under the imposition of the transverse gauge condition $\aa_\n \hh^{\m\n}=0$, namely 
%%%%%%%%%%%%%%%%
\begin{equation}
\da \hh_{\m\n}\,= -2k\,T_{\mu\nu}\,. \nonumber
\end{equation}
%%%%
The proportionality constant is taken to be $k = - \frac{2 \pi C}{c^4}$ and the energy-momen-tum tensor is $T_{\m\n}\,=\,\r\,u_\m u_\n$, with $\r$ the electric charge density. 

The ansatz which we use here is the same as in the subsection (\ref{E2})
%%%%%%
\beq 
\tilde h_{00}=\frac{\Phi}{c^2}\,, \qquad \tilde h_{0i}=\frac{A_i}{c^2}\,,
\qquad \tilde h_{ij}=- \frac{\F}{c^2}\,\h_{ij}\,+\,\frac{2}{c^4}\,d_{ij}\,. \label{case9}
\eeq
%%%%%%
Nevertheless, here $\F$ and $\vec{A}$ are recognized as the true electromagnetic potentials, while the potential $d_{ij}$ denotes the extra degrees of freedom and it should obviously be related with the new fields. In the above ansatz the potentials have the same magnitude and we can also define the electromagnetic 4-potential as $A^\m=\left(\,\F\,,\,\vec{A}\,\right)$.

The transverse gauge condition  gives:

\begin{itemize}

\item For $\m=0$,  the Lorentz gauge condition:
%%%%%%%%
\begin{equation}\label{lor9}
\aa_t \frac{\Phi}{c^3} +\aa_i\,\frac{A^i}{c^2}\,=\,0 \4 or \4 A^{\,\m}_{\,\, ,\m}  =0 . 
\end{equation}
%%%%%%%%

\item And for $\m=i$, an equation which relates the electric field with the divergence of the tensor potential $d_{ij}$:
%%%%%%%%
\begin{equation}\label{add9}
\aa_t \frac{A^i}{c}  -\aa^i\F\,=\,E^i\,=\,\frac{2}{c^2} \,\aa_jd^{ij} .
\end{equation}
%%%%%%%%%
  
\end{itemize}  
Equation (\ref{lor9}) has the well-known form of the Lorentz gauge condition. From Eq. (\ref{add9}) we get an additional constraint for the tensor potential $d_{ij}$; it must be related to the electric field. In the last equation we can see the usefulness of the potentials $d_{ij}$: the  problematic constraints for the electromagnetism  are now constraints regarding the additional fields.  

The field equations take the form:

\begin{itemize}

\item For $\m=0 $ and $\n=0\,$:
\begin{equation}\label{poison9}
\da \frac{\Phi}{c^2} =\, \frac{4 \p C}{c^2}\,\r.
\end{equation}

\item For $\m=0 $ and $\n=i\,$:
\begin{equation}\label{dianysm9}
\da A^i\,=\, \frac{4 \p C}{c}\,j^i.
\end{equation}

\item Finally for $\m=i $ and $\n=j\,$:
\begin{equation}\label{adeq9}
\da \, \left(- \frac{\F}{c^2} \,\h_{ij}\,+\,\frac{2}{c^4}\,d_{ij} \right) \,=\, \frac{4 \p C}{c^4}\,j^i\,u^j\,,
\end{equation}
and if we use Eq. (\ref{poison9}), Eq. (\ref{adeq9}) takes the form:
%%%%%%%
\begin{equation}\label{adeq91}
\da  \,d^{ij} =\, 2\,\p C\left( j^i u^j + \r c^2 \h^{ij} \right).
\end{equation} 

\end{itemize}
%%%%
 
%%
Equations (\ref{poison9}-\ref{dianysm9}) are the correct equations for the electromagnetic potential and along with Eq. (\ref{adeq91}) are the full system of the field equations. 

Now we can define  the electric and   magnetic fields   as usual :
%%%%%%%%%%%%%%%
\beq\label{ebdef9}
\vec{E} \equiv - \,\partial_t \frac{\vec{A}}{c} -
\vec{\nabla} \Phi  \,, 
\qquad \vec{B} \equiv \vec{\nabla} \times \vec{A}\,.
\eeq
%%%%%%%%%%%%%%%
Then one may easily see that Eqs. (\ref{poison9}) and (\ref{dianysm9}) along with Eqs. (\ref{ebdef9}) reduce to the four Maxwell  equations for the  electromagnetic fields:
%%%%%%%%%%%%%%
\beq
\vec{\nabla} \cdot \vec{E} = 4\pi C \rho\,, \qquad 
\vec{\nabla} \times \vec{B}= \,\partial_t \vec{E} + 
\frac{4\pi C}{c}\,\vec{j}\,, \label{FinalGEM91}
\eeq
and
\begin{equation}\label{FinalGEM92}
\vec{\nabla} \times \vec{E}= \,\partial_t \vec{B}, \4\4 \vec{\nabla} \cdot \vec{B} =0.
\end{equation}
The new fields can be defined as:
%%%
\begin{equation}
M_{ij}=-2\, \aa_t \,d_{ij}
\end{equation}
%%%%%
and 
%%%%%%
\begin{equation}
K_{ijk}=-\, \left(\aa_j\,d_{ik}+\aa_k\,d_{ij}-\aa_i\,d_{jk}\right).
\end{equation}
%%%%%%%%
The above definitions will become clear when we  give the expression for the Lorentz force.

The line element for this ansatz is:
\begin{align}\label{line-element-case9}
  ds^2= \,c^2\left(1+\frac{2\Phi}{c^2}\right) dt^2-\frac{2}{c}\,(\vec{A} \cdot d\vec{x})\,dt  
 + \left(\h_{ij}+\frac{2}{c^4}\,d_{ij}\,\right) dx^i dx^j\,.
\end{align}
We can now  give the equations of motion of a test particle moving in the background (\ref{line-element-case9}). Keeping all the corrections the Christoffel symbols for the line element (\ref{line-element-case9}) are:
%%%%
\begin{align}\label{chriscas9}
\G^i_{00} &= \frac{1}{c^2}\left( \aa_0A^i -\aa^i\F \right), \\[2mm] 
%%%%
\G^i_{0j} &= \frac{1}{2} \h^{ik}\left(\frac{1}{c^2}\,F_{jk}+\frac{2}{c^5}\, \aa_t \, d_{jk}\,\right) \,,  \\[2mm]
%%%%
\G^i_{kj} &= \frac{1}{c^4}\left(\aa_k\,d_j^{\,\,\,i} \,+\,\aa_j\,d_k^{\,\,\,i} \,\,-\, \aa^i\,d_{jk}\, \right),
\end{align}
%%%%%%%%%%%%
 Substituting these into the geodesics equation (\ref{geodesics1-GEM51}), we
obtain the equations of motion
%%%%%%%%%%%%%%%
\begin{align}\label{lortzfullc9}
\frac{m}{q}\ddot{x}^{\,i} =\,   \, E^i +\frac{1}{c} F^{ij}u_j \, \,-\frac{1}{c^4}\,  \left( 2\,u^j\, \aa_t d^i_{\,\,\,j}    +2\,u^iu^k\,\aa_k\,d^i_{\,\,\,j} -\,u^ju^k\,\aa^i\,d_{jk} \right).
\end{align}
If we use the definitions for the new fields, the above equation   takes the form:
%%%%%
\begin{equation}
\frac{m}{q}\ddot{x}^{\,i} =\,   \, E^i +\frac{1}{c} F^{ij}u_j \, +\frac{1}{c^4} M^{ij}\,u_j + \frac{1}{c^4} K^i_{\,\,jk}u^ju^k
\end{equation}
%%%%%%%
As we can see, the above equation contains additional corrections; nevertheless,  due to the presence of the $1/c^4$ factor, even at high velocities, it can take the form of the Lorentz force law:
\begin{equation}\label{lortzc9}
\frac{m}{q}\ddot{x}^{\,i}= \, E^i + F^{ij}u_j \,,
\end{equation}
or in a more familiar vector notation:
\begin{equation}\label{lortzvec9}
m\ddot{\vec{x}}\,=\, q\left(\vec{E}+\vec{u}\times \vec{B} \right).
\end{equation}
In this section we have showed that, while the use of the additional degrees of freedom of the theory   has as a result the appearance of new fields, the contribution of these new fields in observable quantities is negligible. The only field that remains even at high velocities is the electromagnetic field. This means that the theory which we study here   could be considered as a realistic theory of electromagnetism \cite{bako}. 

\clearpage
\thispagestyle{empty}

\chapter{Seeking for invariants }\label{6}

\section{The theoretical framework}\label{I}

In this chapter we  focus on the invariant quantities we can get from the General Relativity in the context of gravitoelectromagnetism. Actually, we are interested in invariants similar to the ones appearing in the true Electromagnetism i.e ($B^2-E^2$ and $\vec{E}\cdot \vec{B}$). The first one comes from the inner product  of the electromagnetic tensor
%%%
\begin{equation}
F^{\m\n}F_{\m\n}= 2\,(B^2 -E^2), 
\end{equation}
and appears in the expression of the Lagrangian of the electromagnetic field
%%%
\begin{equation}
\mathcal{L}_{ED}= -\frac{1}{4}F^{\m\n}F_{\m\n}+J^\m A_\m,
\end{equation}
%%%%%
and also, in the expression of the stress-energy tensor of electromagnetism
\begin{equation}
T^{\m\n}= F^{\m\a}F^\n_{\2\a}-\frac{1}{4}g^{\m\n}F^{\a\b}F_{\a\b}.
\end{equation}
The second one comes from the product of the electromagnetic tensor with its dual tensor
\begin{equation}
\tilde{F}^{\m\n} F_{\m\n}=-4 \,\vec{E}\cdot \vec{B},
\end{equation}
%%%
and  is obviously a pseudoscalar since the dual tensor of the electromagnetic tensor is defined as
\begin{equation}
\tilde{F}^{\m\n} = \frac{1}{2}\e^{\m\n\a\b}F_{\a\b}.
\end{equation}
%%%
In the above equations, for simplicity,  we have set $c=\e_0=\m_0=1$.

 We already know that we can construct invariants similar to the electromagnetic ones  from the Weyl tensor \cite{Costa}, but from a different approach of gravitoelectromagnetism. Here we  try to get invariants similar to the electromagnetic ones from the ``linearised" theory that we have used in the previous chapters.

We know that we can get many invariants from the General Relativity, such as the scalar curvature 
\begin{equation}
R^{\a \b\m\n} R_{\a\b\m\n},
\end{equation}
 the Ricci scalar
\begin{equation}
R=g_{\m\n}R^{\m\n},
\end{equation}
or any other scalar that we can construct using the Riemann  and Ricci  tensors. The problem is that the above tensors contain second derivatives of the metric tensor. More specificaly, in the definition of the fields  only first derivatives of the metric tensor are involved, and,  as a result, we can not connect the fields with the gravitational scalars which we have mentioned above. Therefore, the only way to get invariants similar to the electromagnetic ones is to define a new gravitational  tensor, which  is going to contain only first derivatives of the metric tensor, and to construct scalars from this tensor.

In the weak field approximation, we can easily construct a tensor which contains only first derivatives of the metric tensor from the Einstein's   tensor
%%%%%%%%%%%%%%%%%
\begin{equation} 
G_{\m\n}=\frac{1}{2}\left({\hh^{\a}}_{\2\m,\n\a}+{\hh^{\a}}_{\2\n,\m\a}-
\da \hh_{\m\n}-\h_{\m\n}\,{\hh^{\a\b}}_{\4,\a\b}\right).  \nonumber 
\end{equation}
%%%%%%%%%%%%%%%%%%%
If we ``pull" a derivative out, the Einstein's tensor takes the form
%%%%
\begin{equation}\label{einnew}
G_{\m\n}=\frac{1}{2}\aa^\a F_{\a\m\n},
\end{equation}
%%%%%%%
where the tensor $F_{\a\m\n}$ is defined as
%%%%%
\begin{equation}\label{fdef}
F_{\a\m\n}= \aa_\m \hh_{\a\n} + \aa_\n \hh_{\a\m} - \aa_\a \hh_{\m\n} - \h_{\m\n} \aa^\b \hh_{\a\b},
\end{equation}
%%%
and   is obviously symmetric in the last two indices.
Using the $F_{\a\m\n}$ tensor the field equations (\ref{field_eqs_full}) take a form very close to the form of the field equations of electrodynamics   ($\aa_\m F^{\m\n} = \frac{4 \p C}{c^2} \, J^\n  $)
%%%%%%%%
\begin{equation}\label{fieldten}
\aa^\a F_{\a\m\n}=2k T_{\m\n}.
\end{equation}
%%%
We, thus, conclude that we can use the tensor $F_{\a\m\n}$ in order to construct a number of scalars in which the expressions of the fields should appear. However, having in mind the analogy between the field equations of the two theories, we are expecting that the scalar
%%%%%
\begin{equation}\label{fscaldef}
F=F^{\a\m\n}F_{\a\m\n}
\end{equation}
%%%%%%%
will contain a term similar to the $B^2-E^2$ term of the true electromagnetism. In the following sections we are going to calculate the scalar $F$ for the two ansatzes of chapter \ref{4} and to see that   for both ansatzes we can get the requested term.

Our work on the derivation of the second electromagnetic scalar ($\vec{E}\cdot \vec{B}$)  is still in progress \cite{bako}. We have not yet managed to construct a scalar which   contains a term similar to the $\vec{E}\cdot \vec{B}$ term of the true electromagnetism,  but we are expecting that a scalar of the form 
%%%%%
\begin{equation}\label{scal2}
  \tilde{F}^{\a\m\n}F_{\a\m\n} 
\end{equation}
will contain such a term. In the above expression we have defined the dual tensor of $F_{\a\m\n}$ as
%%%%%
\begin{equation}\label{dualdef}
\tilde{F}^{\a\m}_{\4\,\,\,\n}=\e^{\a\m\l\r}F_{\l\r\n}.
\end{equation}
The dual tensor is antisymmetric in the first two indices  and, if the analogy with the true electromagnetism is still persistent, we are expecting that the  expression 
%%%%
\begin{equation}
\aa^\a\tilde{F}_{\a\m\n}=0
\end{equation}
%%%
will contain the other two ``Maxwell-like" equations; the ones that we have used to get straight from the definitions of the fields and not from a fundamental equation. 

\section{The scalar F for the general form of the traditional ansatz}\label{J}

In this section we   calculate the F scalar for the general form of the traditional ansatz 
%%%%%%%%%%%%
\beq 
\tilde h_{00}=\frac{4\Phi}{c^2}\,, \qquad \tilde h_{0i}=\frac{2A_i}{c^2}\,,
\qquad \tilde h_{ij}=\frac{2\l}{c^4}\,\h_{ij}\,+\,\frac{2}{c^4}\,d_{ij}\,. \nonumber
\eeq
%%%%%%%%%%%
%
The components of the $F_{\a\m\n}$ tensor for the above ansatz are
%%%%%%
\begin{align}  
F_{000}=&-\frac{2}{c^2} \aa^i A_i, \label{eq1}\\[2mm]
F_{00i}=&\2\2\frac{4}{c^2} \aa_i \F, \\[2mm]
F_{i00}=&\2\2\frac{4}{c^2}E_i - \frac{2}{c^4}  \left(\aa_i \l + \aa^j d_{ij} \right), \\[2mm] 
F_{ij0}=&- \frac{2}{c^2}F_{ij} + \frac{2}{c^5}\aa_t \left(\l \h_{ij} +d_{ij} \right), \\[2mm]  
F_{0ij}=& - \frac{2}{c^2} \left[\h_{ij} \left(2\aa_0 \F + \aa^k A_k \right) - \left( \aa_i A_j + \aa_j A_i \right)   \right] - \frac{2}{c^5}\aa_t \left(\l \h_{ij} +d_{ij} \right),\\[2mm]
F_{ijk} =& - \frac{2}{c^4} \left[ \aa_i \left(\l \h_{jk} + d_{jk} \right) - \aa_j \left(\l \h_{ik} + d_{ik} \right) - \aa_k \left(\l \h_{ij} + d_{ij} \right) \right] \nonumber, \\[2mm]
\,&- \frac{2}{c^3}\h_{jk} \left[ \aa_t A_i +\frac{1}{c} \aa^l \left(\l \h_{il} + d_{il} \right) \right], \label{eq6}
\end{align}
%%%%%%
where the fields are defined as in Eq. (\ref{EB_GEM0})
%%%%%%%%%%%%%%%
\beq
E_i \equiv \frac{1}{c}\,\partial_t \left(\frac{A_i}{2}\right) -\aa_i \Phi\,, 
\qquad F_{ij} \equiv \aa_i A_j - \aa_j A_i =- 2\e_{ijk}B_k.
\eeq
%%%%%%%%%%%%%%%

Then, the scalar $F=F^{\a\m\n}F_{\a\m\n}$   given in Eq. (\ref{fscaldef}),    can  be written   as
%%%%
\begin{equation}\label{fexp}
F=F^{000}F_{000}+2F^{00i}F_{00i}+F^{0ij}F_{0ij}+F^{i00}F_{i00}+2F^{ij0}F_{ij0}+F^{ijk}F_{ijk}.
\end{equation}
The dominant terms of order  $\mathcal{O}(c^{-4})$    are 
%%%%%
\begin{align}\label{fscac4}
c^4 F = & - 16 E^2 +64 B^2 +4 \left(\aa^iA_i\right)^2+32 \left(\aa_i \F \right)^2 +4 \left( \aa_i A_j +\aa_j A_i \right)^2 \nonumber \\[2mm]
\,& -16 \left(\aa^iA_i\right) \left(2\aa_0\F + \aa^jA_j \right) + 12 \left(2\aa_0\F + \aa^jA_j \right)^2.
\end{align}
We do not give the full expression of the scalar $F$ here because the terms which we want are appearing only in the lowest order. However,  it is easy for someone  to calculate the full expression for the scalar $F$ from Eq. (\ref{fexp}) by using the components of the  $F_{\a\m\n}$ tensor  (\ref{eq1}-\ref{eq6}). We should also mention that in the above calculations we do not use any gauge condition. If we use the transverse gauge condition $\aa^\n \hh_{\m\n}=0$, the last term in the definition of the $F_{\a\m\n}$, Eq. (\ref{fdef}),   disappears  and as a result the last two terms of the scalar $F$, Eq. (\ref{fscac4}), also disappear. 

Concluding, we have managed to construct a scalar combination in terms of gravitational quantities in which a term  like the electromagnetic one  $(B^2-E^2)$ appears. However, this term is followed by additional terms due to the tensorial structure of the theory.

\section{The scalar F for the general form of the alternative ansatz}\label{K}

In this section we   calculate the F scalar for the general form of the alternative ansatz  
%%%%%%%%%%%%
%%%%%%
\beq 
\tilde h_{00}=\frac{\Phi}{c^2}-\frac{3\l}{c^4}\,, \qquad \tilde h_{0i}=\frac{A_i}{c}\,,
\qquad \tilde h_{ij}=-\left(\frac{\F}{c^2}+\frac{\l}{c^4}\right)\,\h_{ij}\,+\,\frac{2}{c^4}\,d_{ij}\,. \nonumber 
\eeq
%%%%%%
The components of the $F_{\a\m\n}$ tensor for the above ansatz are
%%%%%%
\begin{align}
F_{000}=& -\frac{1}{c^2} \aa^i A_i, \label{eqq1} \\[2mm]
F_{00i}=& \4 \frac{1}{c^2}\aa_i \F - \frac{3}{c^4}\aa_i \l, \\[2mm]
F_{i00}=& \4 \frac{1}{c^2}\left(4E_i +\aa_i \F \right) + \frac{2}{c^4} \aa^j \left(\l \h_{ij} + d_{ij} \right), \\[2mm]
F_{ij0}=& -\frac{1}{c^2}F_{ij} - \frac{1}{c^3}\aa_t \left[\left(\F +\frac{1}{c^2}\l \right)\h_{ij} -\frac{2}{c^2} d_{ij} \right], \\[2mm]
F_{0ij}=& \4 \frac{1}{c^2} \left[  \left( \aa_iA_j +\aa_jA_i \right) -  \h_{ij} \left(\aa_0\left(\F -\frac{3}{c^2}\l\right)+\aa^lA_l \right) \right] \nonumber \\[2mm]
\,& + \frac{1}{c^3}\aa_t \left[\left(\F +\frac{1}{c^2}\l \right)\h_{ij} -\frac{2}{c^2} d_{ij} \right], \\[2mm]
F_{ijk}=& -\frac{1}{c^2} \left(\h_{ij}\aa_k \F +\h_{ik}\aa_j \F -2\h_{jk} \aa_i \F \right) +\frac{1}{c^4}\h_{jk}\aa^l\left( \l\h_{il}-2d_{il}\right) \nonumber \\[2mm]
\,&-\frac{1}{c^4} \left[ \aa_k \left(\l \h_{ij} - 2d_{ij} \right) + \aa_j \left(\l \h_{ik} -2 d_{ik} \right) - \aa_i \left(\l \h_{jk} -2 d_{jk} \right) \right] \nonumber \\[2mm]
\,& - \frac{1}{c^3}\h_{jk}   \aa_tA_i,   
\end{align}
where the fields are defined as in Eq. (\ref{ebdef})
\begin{equation}
E_i \equiv \frac{1}{4} \left( \frac{1}{c}\,\partial_t  A_i   -\aa_i  \Phi \right)\,, 
\qquad F_{ij} \equiv \aa_i A_j - \aa_j A_i =- 4\e_{ijk}B_k.
\end{equation}
However, as we have already seen in section (\ref{E}), the above fields are recognized as the gravitoelectromagnetic fields, only if the scalar potential $\l$  vanishes.

Finally the lowest order  $\mathcal{O}(c^{-4})$  terms of the scalar $F$ are 
%%%%%%%
\begin{align}\label{fsca23}
c^4 F =& -16E^2 + 32 B^2 + \left(\aa^iA_i\right)^2 + 3\left((\aa_i \F \right)^2 +8 E_i \aa^i \F \nonumber \\[2mm]
\,& +\left[ \left(\aa_iAj+\aa_jA_i\right) -\h_{ij} \left(\aa_0\left(\F - \frac{3}{c^2}\l \right) +\aa^lA_l\right) \right]^2 \nonumber \\[2mm]
\,& +\left( \h_{ij} \aa_k \F + \h_{ik}\aa_j \F -2 \h_{jk} \aa_i \F \right)^2.
\end{align}
Here, as in the previous section, we give the dominant  terms of the scalar $F$ without the imposition of a gauge condition. If we use the transverse gauge condition, the term
$
\left[\aa_0\left(\F - \frac{3}{c^2}\l \right) +\aa^lA_l\right]
$
%%%
 in   Eq. (\ref{fsca23})  vanishes. In this ansatz the  scalar $F$   also contains extra terms. In the case of gravitoelectromagnetism the extra terms do not create any problem; they  just show  us where the analogy with the true electromagnetism stops. 
However, in  chapter \ref{5} we have used the above ansatz in order to give an alternative description of the true electromagnetism. If we have  alternative descriptions of a theory, they should  all give us the same results. The presence of the additional terms in the scalar $F$ --for the analysis of  chapter \ref{5}--   is then problematic.

%%%%%%%%%%%%%%%%%%%%%%%%%%%%%%%%%%%%%%%%%%%%%%%%%%%%%%%%%%%%%%%%%%%%%%%%%%%%%%%%%%%

\chapter{Conclusions}\label{7}

There is undoubtedly a striking similarity between the gravitational and electromagnetic
forces at classical level. These similarities, which persist even  in the context of the General Theory of Relativity, led to the development of gravitoelectromagnetism. In our days the majority of the gravitoelectromagnetism approaches are split in two large categories. In the first one the analogy between gravitation and electromagnetism is shown from the linearised form of Einstein's field equations, in the weak field approximation, around a rotating massive body. In the second, the analogy is shown from the decomposition of the Weyl tensor into gravitoelectric and gravitomagnetic parts. The main purpose of this thesis was to fully review the first approach of gravitoelectromagnetism, to shed light to particular problematic aspects of the theory and, also, to propose solutions.

In chapter \ref{2} we gave a historical review of gravitoelectromagnetism and also a brief introduction to General Relativity and to the linear approach. We considered this as particularly important in order to give the context in which our work was based and to provide the reader with all the information needed to understand better our project.

In chapter \ref{3} we reviewed the linear approach of gravitoelectromagnetism. In the first section of the chapter, we started from the perturbed Einstein's equations in the linear approximation and, employing the so-called traditional ansatz for the metric perturbations, we showed the analogies between gravity and electromagnetism. However, these analogies stand only if the gravitoelectromagnetic potentials are static; this along with the presence of additional terms in the equations of motion consist the two big problems of gravitoelectromagnetism.  In the second section, in order to investigate further the problems of the traditional ansatz, we did again all the calculations without the imposition of a gauge condition. We showed that the problem of the time dependence of the vector potential is a problem of the ansatz which can not be solved by choosing an alternative gauge condition. The above, along with the presence of extra terms in the Lorentz force law, led us to the search for new ansatzes for the metric perturbations. In the third section, we proposed an alternative ansatz; this ansatz was chosen carefully, in order not to give extra terms in the equations of motion. However, in this ansatz we encountered a serious problem, that of the vanishing of the source. Of course this ansatz  could be used in a vacuum problem, but in gravitoelectromagnetism the presence of the source is necessary. 

In chapter \ref{4} we tried to resolve the problems appearing in gravitoelectromagnetism. In chapter \ref{3} we showed that in order to solve these problems we must try new ansatzes for the metric perturbations. However, from the equation $\aa_j \hh^{ij} \,=\,-\,\aa_t\,A^i\,$,  we understood that the use of the --suppressed-- additional degrees of freedom of the gravitational field could solve some of our problems. In this chapter we used the extra degrees of freedom in both ansatzes. In the first section we tried the most general form of the traditional ansatz. We showed that the condition, which in the previous chapter was leading to a  static vector potential, in this one,  became a condition which connects the vector potential with the additional degrees of freedom of the gravitational field. However, extra terms in the equation of motion were still present. In the second section we did the same with the alternative ansatz, the one  introduced in chapter \ref{3}. This section was split in two subsections. In the first one we considered the most general ansatz. Although  this ansatz did not lead to a valid model of gravitoelectromagnetism, we included it in our analysis because it could be used in other aspects of gravitational problems. In the second we studied a special case of the above ansatz, which turned out to be the best one in the context of gravitoelectromagnetism because it gives both the correct form of the field equations and the equations of motion.
 
In chapter \ref{5}  we  covered a topic we initiated in a previous work of ours.  In this chapter we attempted to describe true electromagnetism using the formalism of General Relativity. This attempt was based on the results of our work regarding gravitoelectromagnetism. In the first section we presented a short introduction of electromagnetism and we gave an alternative expression for the Lorentz force law using geometrical quantities. In the second section we reviewed our previous work.  In the third section we presented some improvements  to our previous work based on the results of chapter \ref{3}. 

Finally, in chapter \ref{6} we covered   the search for invariant quantities similar to the electromagnetic ones, in the linear approach of gravitoelectromagnetism. In the first section of the chapter we presented the theoretical framework  and we defined a gravitoelectromagnetic tensor similar to the electromagnetic field tensor  $F_{\m\n}$. In the following two sections of this chapter we presented  some calculations of scalars constructed by the gravitoelectromagnetic tensor we have defined.  However, this work is still in progress and we only present some of our results. We hope that this work would contribute  to a better understanding of gravitoelectromagnetism and, maybe,  lead to a lagrangian formulation.

\clearpage
\thispagestyle{empty}

%%%%%%%%%%%%%%%%%%%%%%%%%%%%%%%%%%%%%%%%%%%%%%%%%%%%%%%%%%%%%%%%%%%%%% Vivliografia%%%%%%%%%%%%%%%%%%%%%%%%%%%%%%%%%%%%%%%%%%%%%%%%%%%%%%%%%%%%%%%%%%%%%%%%%

\end{document}